\pdfoutput=1
\documentclass[letterpaper, 10pt, conference]{ieeeconf}  

\IEEEoverridecommandlockouts                              
                                                          
\usepackage{graphicx}
\usepackage{caption}
\usepackage{subcaption}
\usepackage{amssymb}
\usepackage{MnSymbol}
\usepackage[dvipsnames]{xcolor}
\usepackage{siunitx}

\newcommand{\preprintswitch}[2]{#2} 
\newcommand{\RR}{\mathbb{R}}

\title{\LARGE \bf
Gone with the Wind: Nonlinear Guidance for Small Fixed-Wing Aircraft in Arbitrarily Strong Windfields}

\author{Luca Furieri$^{\dagger}$, Thomas Stastny$^{\ast}$, Lorenzo Marconi$^{\dagger}$, Roland Siegwart$^{\ast}$, and Igor Gilitschenski$^{\ast}$
\thanks{This work was supported by the European Commission project SHERPA (\#600958) under the 7th Framework Programme.}
\thanks{$^{\dagger}$Laboratory of Automation and Robotics, Department of Electronics, Computer Science, and Systems, University of Bologna, Bologna, Italy.
        {\tt\small luca.furieri@studio.unibo.com, lorenzo.marconi@unibo.it}}%
\thanks{$^{\ast}$Autonomous Systems Lab, Department of Mechanical \& Processing Engineering, Swiss Federal Institute of Technology (ETH Z\"{u}rich), Z\"{u}rich, Switzerland.
       {\tt\small <firstname.lastname>@mavt.ethz.ch}}%
}

\begin{document}

\maketitle
\thispagestyle{plain}
\pagestyle{plain}

\begin{abstract}
The recent years have witnessed increased development of small, autonomous fixed-wing Unmanned Aerial Vehicles (UAVs).
In order to unlock widespread applicability of these platforms, they need to be capable of operating under a variety of environmental conditions.
Due to their small size, low weight, and low speeds, they require the capability of coping with wind speeds that are approaching or even faster than the nominal airspeed.
In this paper we present a principled nonlinear guidance strategy, addressing this problem.
More broadly, we propose a methodology for the high-level control of non-holonomic unicycle-like vehicles in the presence of strong flowfields (e.g. winds, underwater currents) which may outreach the maximum vehicle speed. 
The proposed strategy guarantees convergence to a safe and stable vehicle configuration with respect to the flowfield, while preserving some tracking performance with respect to the target path. 
Evaluations in simulations and a challenging real-world flight experiment in very windy conditions confirm the feasibility of the proposed guidance approach. 
\end{abstract}

\section{INTRODUCTION}
In recent years, the use of small fixed-wing Unmanned Aerial Vehicles (UAVs) has steadily risen in a wide variety of applications due to increasing availability of open-source and user-friendly autopilots, e.g. Pixhawk Autpilot~\cite{px4website}, and low-complexity operability, e.g. hand-launch. Fixed-wing UAVs have particular merit in long-range and/or long-endurance remote sensing applications. Research in ETH Z\"urich's Autonomous Systems Lab (ASL) has focused on Low-Altitude, Long-Endurance (LALE) solar-powered platforms capable of multi-day, payload-equipped flight~\cite{Oettershagen_FSR2016}, already demonstrating the utility of such small platforms in real-life humanitarian applications~\cite{Oettershagen_JFR2016}. UAVs autonomously navigating large areas for long durations will inherently be exposed to a variety of environmental conditions, namely, high winds and gusts. With respect to larger and/or faster aircraft, wind speeds rarely reach a significant ratio of the vehicle airmass-relative speed. Conversely, wind speeds rising close to the vehicle maximum airspeed, and even surpassing it during gusts, is a frequent scenario when dealing with a small-sized, low-speed aircraft.

Usually in aeronautics, windfields are handled as an unknown low-frequency disturbance which may be dealt with either using robust control techniques, e.g. loop-shaping in low-level loops, or simply including integral action within guidance-level loops. In the case of LALE vehicles, maximizing flight time would further require the efficient use of throttle, thus limiting airspeed bandwith. In order to be able to use such systems safely and efficiently in a wide range of missions and different environments, it is necessary to take care of such situations directly at the guidance level of control, explicitly taking into account online wind estimates. 


A standard approach to mitigate the effect of wind on path following tasks is to exploit the measurements of the inertial groundspeed of the aircraft, which inherently includes wind effects, see \cite{Osborne}, \cite{Derek}. Another approach is to take the wind explicitly into account, either by available wind measurements \cite{McGee} \cite{Rysdyk} or by exploiting a disturbance observer, as in \cite{Cunjia}. Another possibility is described in \cite{Brezoescu}, where adaptive backstepping is used to get an estimate for the direction of the wind.

As to wind compensation techniques, a possible approach is vector fields \cite{Derek} \cite{beard2013implementing}. In \cite{Derek}, an approach based on vector fields is used to achieve asymptotic tracking of circular and straight-line paths in the presence of non neglegible persisting wind disturbances: vector fields are proposed for specific curves (e.g. straight lines, circles). This requires switching the commands when the target path is defined as the union of different parts, which makes the algorithm less uniform and its implementation trickier. Tuning of vector fields is also known to be difficult, as highlighted in \cite{beard2013implementing}.

Another popular approach is based on nonlinear guidance. The strategy proposed in \cite{ParkL1}, utilises a \emph{look-ahead} vector for improved tracking of upcoming paths, introducing a predictive effect. The law was extended in \cite{Park3D} to any 3D path in the non-windy case. Great advantages of this law are that it is easy and intuitive to tune, the magnitude of the guidance commands is always upper-bounded, and it has flexibility in the set of feasible initial conditions. 

The main contribution of this paper is a simple, safe, and computationally efficient guidance strategy for navigation in \emph{arbitrarily} strong windfields. To our knowledge, there is no existing guidance law directly considering the case of the windspeed being higher than the airspeed. The provided design strategy relies on the solution provided in \cite{Park3D}  in absence of wind whose choice for the look-ahead vector will be properly modified in order to cope with arbitrarily strong windspeed.  

{\bf Notation.} We shall use the bold notation to denote vectors in $\RR^3$. For a vector ${\bf v} \in \RR^3$, $\hat {\bf v}$ denotes the associated versor and $\|{\bf v}\|$ the euclidean norm. For two vectors ${\bf v}_1, {\bf v}_2 \in \RR^3$, their scalar and cross products are respectively indicated by ${\bf v}_1 \cdot {\bf v}_2 \in \RR$ and  ${\bf v}_1 \times {\bf v}_2 \in \RR^3$. 

\section{PROBLEM DEFINITION}
\label{sec:PROBLEM DEFINITION}

As we wish to extend the results obtained in \cite{Park3D}, it is useful to define the same mathematical framework. To have a better insight, we will clearly define the control problem for each different scenario, and define a state-space nonlinear formulation. This will allow us to state a robust control problem, which will be useful for analysis in future work. 


\subsection{The Frenet-Serret framework for autonomous guidance}
\label{3DPL} 
 The position of the vehicle is denoted by $\mathbf{r}_M$, which is a vector of $\mathbb{R}^3$ expressed with respect to an inertial reference frame denoted by  ${\cal F}_i$ and described by an orthonormal right-hand basis $(\mathbf{{i}}, \mathbf{{j}}, \mathbf{k})$. We assume that $(\mathbf{{i}}, \mathbf{{j}})$ are co-planar with the flight plane, with $\mathbf{k}$ orthogonal to such a plane. The emphasis of the work is on developing a controller able to cope with strong wind. The latter is a vector $\mathbf{w} \in \mathbb{R}^3$ assumed to be constant and to lie on the flight plane, namely with zero component along $\mathbf{k}$. The vectors $\mathbf{v}_G \in \mathbb{R}^3$ and $\mathbf{a}_M \in \mathbb{R}^3$ in the plane $(\mathbf{{i}}, \mathbf{{j}})$ denote the ground speed and acceleration of the vehicle, the dynamics of the latter is described by
\begin{equation}
\label{eq:point-mass}
\mathbf{\dot{r}}_M=\mathbf{v}_G\,, \qquad 
\mathbf{\dot{v}}_G=\mathbf{a}_M\,.
\end{equation}
Considering flight through a moving airmass, $\mathbf{v}_G=\mathbf{v}_M+\mathbf{w}$, in which $\mathbf{v}_M$ is the vehicle airmass-relative speed (or airspeed). Note that, since $\mathbf{w}$ is constant, $\mathbf{\dot{v}}_G = \mathbf{\dot{v}}_M$. The acceleration $\mathbf{a}_M$ represents the control input.

From a geometric viewpoint, the \emph{vehicle path} is defined as the union of each $\mathbf{r}_M(t)$ for every time $t$. 
 At each $t\geq 0$ the vehicle path can be geometrically characterised in terms of  the {\em unit tangent vector}, the {\em actual orientation}, the {\em tangential acceleration}, the {\em normal acceleration}, the {\em tangential acceleration}, the {\em unit normal vector} and the {\em curvature} of the vehicle path respectively defined as   
\begin{equation}\label{GeometricPathVehicle}
\begin{aligned}
&\mathbf{\hat{T}}_G(t):=\frac{\mathbf{v}_G(t)}{\|\mathbf{v}_G(t)\|}\,, \qquad
\mathbf{\hat{T}}_M(t):=\frac{\mathbf{v}_M(t)}{\|\mathbf{v}_M(t)\|}\,,\\
&{\mathbf{a}_M^T (t)} := (\mathbf{a}_M(t) \cdot  \mathbf{\hat{T}}_M(t)) \mathbf{\hat{T}}_M(t)\,,\\
&\qquad \qquad {\mathbf{a}_M^N (t)} := (\mathbf{\hat{T}}_M(t) \times \mathbf{a}_M(t)) \times  \mathbf{\hat{T}}_M(t) \,,\\
&\mathbf{\hat{N}}_M(t):=\frac{\mathbf{a}_M^N(t)}{\|\mathbf{a}_M^N(t)\|}\,,\quad
k_M(t):=\frac{\|\mathbf{a}_M(t)\|}{\|\mathbf{v}_G(t)\|^2}\,.
\end{aligned}
\end{equation}
We observe that the unit normal vector is defined only for values of the acceleration such that $\|\mathbf{a}_M^N(t)\| \neq 0$.  Furthermore, all the previous 
vectors lie in the plane  $(\mathbf{{i}}, \mathbf{{j}})$.
Having in mind the application to fixed-wing UAVs, we will consider the vehicle to be unicycle-like, i.e. its speed norm $\|\mathbf{v}_M\|$ will remain unchanged in time and it will be then guided through normal acceleration commands ${\mathbf{a}_M^N}$. In other words, the control law for $\mathbf{a}_M$ will be chosen in such a way that ${\mathbf{a}_M^T}(t) \equiv 0$. According to this, and by bearing in mind (\ref{GeometricPathVehicle}), (\ref{eq:point-mass}) can be rewritten as
\begin{equation}
\label{eq:point-mass_normal-acc}
\mathbf{\dot{r}}_M(t)=v_M^\star \mathbf{\hat{T}}_M(t)+\mathbf{w}(t), \quad 
v_M^\star \mathbf{\dot{\hat{T}}}_M(t)={\mathbf{a}_N^M(t) }\\
\end{equation}
in which $v_M^\star$ denotes the (constant) value of $\|\mathbf{v}_M\|$.

Inspired by \cite{Park3D}, the {\em desired (planar) path} is a continuously differentiable  space curve in the plane spanned by $(\mathbf{{i}}, \mathbf{{j}})$ 
represented by ${\mathbf{p}(l)}$, $l \in \mathbb{R}$, with associated a {\em Frenet-Serret} frame composed of three orthonormal vectors $(\mathbf{\hat{T}}_p(l),
\mathbf{\hat{N}}_p(l), \mathbf{\hat{B}}_p(l) )$, a curvature $\kappa_p(l)$ and a torsion $\tau_p(\ell)$. In the following we let $s \in \mathbb{R}$ the arc length along the curve $p(\cdot)$ defined as 
\[
 s(l) = \int_{l_0}^l \|  {d \mathbf{p}(\ell) \over d\ell} \| d \ell\,.
\] 
The desired path is thus endowed with the {\em Frenet-Serret} dynamics given by
\begin{equation}\label{F-Ss}
 \left( \begin{array}{c}
 {\mathbf{\hat{T}}_p'(s)}\\
  {\mathbf{\hat{N}}_p'(s) }\\
 {\mathbf{\hat{B}}_p'(s) }\\
 \end{array}
 \right ) =  \left ( \begin{array}{ccc}
 0 & \kappa_p(s) & 0\\
  -\kappa_p(s) & 0 & \tau_p(s)\\
 0 & - \tau_p(s) & 0 
 \end{array} \right )
  \left( \begin{array}{c}
 { \mathbf{\hat{T}}_p(s) }\\
 {  \mathbf{\hat{N}}_p(s) }\\
 { \mathbf{\hat{B}}_p(s) }
 \end{array}\right )
\end{equation}
in which we used the notation $\left(\cdot\right)'$ to denote the derivative with respect to $s$. 
As in \cite{Park3D}, we define the ``footprint" of $\mathbf{r}_M$ on $\mathbf{p}$ at time $t$ as the closest point of $\mathbf{r}_M(t)$ on $\mathbf{p}(l)$ defined as 
\[
 {\mathbf{r}_P}(s(t)):=\text{arg}\underset{\mathbf{r} \in \mathbf{p}}{\text{ min}}\|\mathbf{r}_M(t)-\mathbf{r}\|\,.
\]  
The point $P$ on the desired path is identified by $l_P$, which is the value of the curve parameter $l$ at the closest projection. The unit tangent vector,  the unit normal vector, the unit binormal vector,  the curvature and the torsion of the desired path at the point $P$ will be indicated in the following as 
$\mathbf{\hat{T}}_P:=\mathbf{\hat{T}}_p(l_P)$, $\mathbf{\hat{N}}_P:=\mathbf{\hat{N}}_p(l_P)$, $\mathbf{\hat{B}}_P:=\mathbf{\hat{B}}_p(l_P)$, $\kappa_P:=\kappa_p(l_P)$ and $\tau_P:=\tau_p(l_P)$. They are all functions of time through $s(t)$.
 By bearing in mind (\ref{F-Ss}), it turns out that the vehicle dynamics induce  a {\em Frenet-Serret} dynamics on the desired path which is given by 
 \begin{equation}\label{F-St}
 \left( \begin{array}{c}
 {\mathbf{\dot {\hat T}}}_P(t) \\
 {\mathbf{\dot {\hat N}}}_P(t)\\
 {\mathbf{\dot {\hat B}}}_P(t)
 \end{array}
 \right ) = \dot s(t) \left ( \begin{array}{ccc}
 0 & \kappa_p(t) & 0\\
  -\kappa_p(t) & 0 & \tau_p(t)\\
 0 & - \tau_p(t) & 0 
 \end{array} \right )
  \left( \begin{array}{c}
 { \mathbf{\hat{T}}_P(t) }\\
 {  \mathbf{\hat{N}}_P(t) }\\
 { \mathbf{\hat{B}}_P(t) }
 \end{array}\right )
 \end{equation}
 in which $ \dot s(t)$ can be easily computed as (see Lemma 1 and Appendix B in \cite{Park3D}).
 
 \[
  \dot s(t) = { (v_M^\star \mathbf{\hat{T}}_M(t)+\mathbf{w}) \cdot   \mathbf{\hat{T}}_P(t)  \over 1 +\kappa_P(t) [(\mathbf{r}_P(t) - \mathbf{r}_M(t))\cdot \mathbf{\hat{N}}_P(t)]}\,.
 \]
 The (ideal) desired  control objective is to asymptotically steer the position of the vehicle $\mathbf{r}_M(t)$ to the  footprint ${\mathbf{r}_P}(s(t))$ by also aligning the unitary tangent vectors $\mathbf{\hat{T}}_G(t)$ and $\mathbf{\hat{T}}_P(t)$ and their curvature.  To this end it is worth introducing an error $\mathbf{e}(t)$ defined as 
 \[
 \mathbf{e}(t):=\mathbf{r}_P(t)-\mathbf{r}_M(t)
 \]
 and to rewrite the relevant dynamics in error coordinates. In this respect, by considering the system dynamics (\ref{eq:point-mass}), the Frenet-Serret dynamics (\ref{F-St}), it is simple to obtain (for compactness we omit the arguments $t$) 
\begin{equation}
\label{eq:nlsys_wind}
\hspace*{-3mm}\begin{array}{rcl}
\mathbf{\dot e} &=& 
 \displaystyle - \left ( \mathbf{v}_G  \cdot \mathbf{\hat{T}}_P \right ) \left( 
 \frac{\kappa_P(\mathbf{e} \cdot \mathbf{\hat{N}}_P) }{1+\kappa_P(\mathbf{e} \cdot \mathbf{\hat{N}}_P)} \mathbf{\hat{T}}_P + \mathbf{\hat{N}}_P
  \right )\\
\mathbf{\dot{\hat{T}}}_P&=&\displaystyle \frac{\kappa_P(\mathbf{v_G} \cdot \mathbf{\hat{T}}_P)}{1+\kappa_P(\mathbf{e} \cdot \mathbf{\hat{N}}_P)}\mathbf{\hat{N}}_P\\
\mathbf{\dot{\hat{N}}}_P&=&
\displaystyle \frac{(\mathbf{v}_G \cdot \mathbf{\hat{T}}_P)}{1+\kappa_P(\mathbf{e} \cdot \mathbf{\hat{N}}_P)}\left ( \tau_P \mathbf{\hat{B}}_P-\kappa_P\mathbf{\hat{T}}_P\right )\\
\mathbf{\dot{\hat{B}}}_P&=&
\displaystyle \frac{- \tau_P(\mathbf{v}_G \cdot \mathbf{\hat{T}}_P)}{1+\kappa_P(\mathbf{e} \cdot \mathbf{\hat{N}}_P)}\mathbf{\hat{N}}_P\\
v_M^\star \mathbf{\dot{\hat{T}}}_M &=&{\mathbf{a}_M^N}
\end{array}
\end{equation}
with the ground speed $\mathbf{v}_G$ that is a function of  $\mathbf{\hat{T}}_M$ and $\mathbf{w}$ according to 
\[
\mathbf{v}_G = v_M^\star \mathbf{\hat{T}}_M+\mathbf{w}\,.
\]
This is a system with state $(\mathbf{e}, \mathbf{{\hat{T}}}_P, \mathbf{{\hat{N}}}_P, \mathbf{{\hat{B}}}_P, \mathbf{\hat{T}}_M)$ with control input $\mathbf{a}_M$  (to be chosen so that $\mathbf{a}_M^T\equiv 0$) subject to the wind disturbance $\mathbf{w}$. Note that for planar paths, $\tau_P=0$.
			
In the paper, similarly to \cite{Park3D},  the acceleration command will be chosen as 
\begin{equation}
\label{eq:normal_acc_cmd}
{\mathbf{a}_M^N} =(\mathbf{v}_M \times {\bf u}) \times \mathbf{v}_M
\end{equation}
with ${\bf u} \in \RR^3$ an auxiliary input to be chosen.  Note that this choice guarantees that ${\mathbf{a}_M^T}(t) \equiv 0$ for all possible choices of $\bf u$. The degree-of-freedom for the problem is then the choice of the control input $\mathbf{u}$ to accomplish control goals. 


\noindent Motivated by \cite{Missile}, the choice of $\bf u$ presented in the paper relies on the so-called \emph{look-ahead vector}, denoted by $\mathbf{\hat{L}}$, which represents the {\em desired groundspeed direction} for the vehicle. The latter will be taken as a function of the  system state and of the wind, according to the objective conditions in which the vehicle operates.

\subsection{Feasibility Cone and Control Objective Formulation}
\label{sub:Control Objective}
 Although the ideal control objective is to steer the error $\mathbf{e}(t)$ asymptotically to zero by also aligning the unitary tangent vectors $\mathbf{\hat{T}}_G(t)$ and $\mathbf{\hat{T}}_P(t)$ and their curvature, the presence of ``strong" wind could make this ideal objective infeasible. For this reason we set two objectives that will be targeted according to the wind conditions.  
 
 {\bf Ideal Tracking Objective.} Ideally, the control input $\bf u$ must be chosen so that the following asymptotic objective is fulfilled
  \begin{equation}
\label{eq:goals:path-tracking_wind}
\begin{cases}
\underset{t->\infty}{\lim}\mathbf{e}(t)=0\\
\underset{t->\infty}{\lim}(\mathbf{\hat{T}_G}(t)- \mathbf{\hat{T}}_P(t))=0\\
\underset{t->\infty}{\lim}(\displaystyle \frac{d\mathbf{\hat{T}_G}(t)}{dt}- \frac{d\mathbf{\hat{T}}_P(t)}{dt})=0
\end{cases}
\end{equation}
namely position, ground speed orientation, and ground speed curvature of the vehicle converge to the path ones.

{\bf Safety Objective.} When strong wind does not allow to achieve the ideal objective, the degraded safety objective consists of controlling the vehicle in such a way that the vehicle acceleration is asymptotically set to zero, the  groundspeed value is asymptotically minimised (by pointing the nose the vehicle against wind) and the vehicle nose asymptotically points to $\mathbf{P}$, namely 
\begin{equation}
\label{eq:goals_finitelength}
\begin{cases}
\underset{t\rightarrow \infty}{\lim} {\mathbf{{a}}_M^N}(t) =0\\
\underset{t\rightarrow \infty}{\lim} \mathbf{\hat{T}}_M(t)={-\hat {\bf w}}\\
\underset{t\rightarrow \infty}{\lim} \hat {\bf e}(t)  = -\hat {\bf w}\,.
\end{cases}
\end{equation}
The targeted configuration, in particular, is the one in which the vehicle goes away with the wind, by minimising the groundspeed (safety objective), and minimising the distance to the closest point on the path. Note that this objective makes sense for \emph{finite-length} paths: the infinite-length linear path case is briefly discussed in \cite{arxiv}.\\
 Ideal or degraded objectives are set according to the fulfilment of  a ``feasibility condition" by the look ahead vector. More precisely, with $w^\star:= \|{\bf w}\|$ the wind strength, let $\beta$ be defined as
  \begin{equation}
\begin{array}{rcl}
\beta &:=& \left \{
 \begin{array}{ll}
 \displaystyle  \arcsin{\frac{v_M^\star}{w^\star}}& \quad  w^\star \geq v_M^\star\\
 \pi & \quad w^\star<v_M^\star\,.
 \end{array}
 \right .
 \end{array}
\end{equation}
  Then, we define the ``feasibility cone" $\cal C$ as the cone with apex centred at the vehicle position ${\bf r}_M$, main axis given by $\bf w$ and with aperture angle $2 \beta$ (see Figure \ref{ink_base_higher_winds}).
Notice that this becomes the entire plane when $w^\star<v_M^\star$, i.e. $\beta=\pi$. All desired groundspeed vectors that lie in the cone can be indeed enforced by appropriately choosing the control input $\bf u$. This fact, and the fact that the look ahead vector represents the desired groundspeed direction for the vehicle, motivates the fact of considering the ideal tracking objective feasible at a certain time $t$ if it's possible to shape the look ahead vector ${\bf \hat{L}}(t)$ so that it lies in $\cal C$.
More specifically, if
\begin{equation}
\label{eq:feasibility}
\lambda=\arccos{\hat {\bf w}\cdot \mathbf{\hat{L}}(t) }<\beta\,.
\end{equation}
  Otherwise, the ideal tracking objective is said infeasible at time $t$. 
   \noindent The control objectives are set consequently, and we  shape the control input separately for each subcase according to the following scheme:
\begin{equation}
\label{eq:full_u}
\mathbf{u}=
\begin{cases}
\mathbf{u}_{\text{slow}}\qquad w^\star \leq v_M^\star\\
\mathbf{u}_{\text{fast,1}} \qquad w^\star > v_M^\star , ~\lambda\leq\beta\\
\mathbf{u}_{\text{fast,2}} \qquad w^\star > v_M^\star, ~\lambda>\beta
\end{cases}
\end{equation}
In sections \ref{sec:SLOWERWIND}, \ref{sec:higher-winds}, we show how to design the control input $\mathbf{u}$ as in (\ref{eq:full_u}) such that if the ideal tracking objective is feasible then (\ref{eq:goals:path-tracking_wind}) is achieved, otherwise the Safety Objective is enforced.

\subsection{The Nominal Solution in Absence of Wind in \cite{Park3D}}
\label{sub:Nominal Solution}
\begin{figure}[htbp]
\centering
\includegraphics[width=0.36\textwidth]{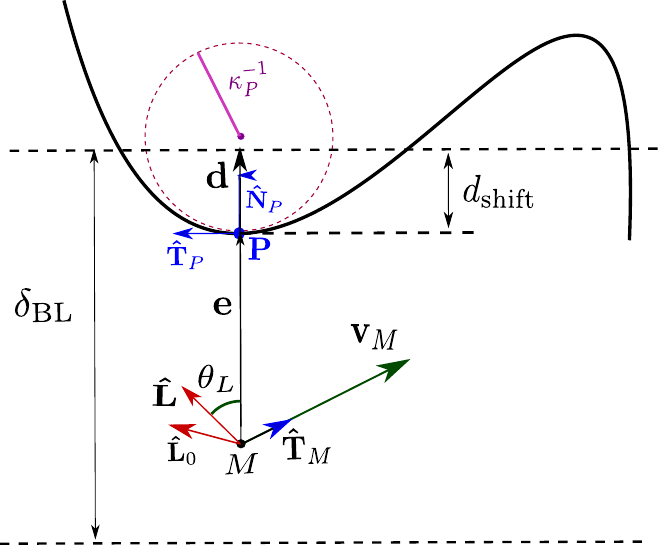}
\caption{Sketch of the nominal solution of \cite{Park3D}}
\label{fig:ink_Park_sketch}
\end{figure}

In this section we briefly present the solution chosen in \cite{Park3D} for the look-ahead vector in absence of wind, as it represents the basis for developing the windy solution. A graphical sketch showing the notation is provided in Figure \ref{fig:ink_Park_sketch}.
\noindent The authors in \cite{Park3D} proposed the control law
\begin{equation}
\mathbf{u}=k\mathbf{\hat{L}}
\end{equation}
in which $k$ is a design parameter chosen so that $k>\underset{P \in p(l)}{\text{max}}k_P$ and 
$\mathbf{\hat{L}}$ is the look-ahead vector chosen as 
\begin{equation}
\label{eq:L}
\mathbf{\hat{L}}=\cos{(\theta_L({\|\mathbf{d}\|}))}\mathbf{\hat{d}}+\sin{(\theta_L({\|\mathbf{d}\|}))}\mathbf{\hat{T}}_P
\end{equation}
where $\mathbf{d}=\mathbf{e}+d_{\text{shift}}\mathbf{\hat{N}}_P=(\|\mathbf{e}\|+d_{\text{shift}})\mathbf{\hat{N}}_P$ is the \emph{radially shifted distance}, $\theta_L(\|\mathbf{d}\|)$ is the function 
\begin{equation}
\label{eq:choice_L}
\theta_L(\|\mathbf{d}\|)=\frac{\pi}{2}\sqrt{1-\text{sat}(\frac{\|\mathbf{d}\|}{\delta_{BL}})}
\end{equation}
in which $\delta_{BL}$ is  a \emph{boundary layer parameter} and the parameter $d_{\text{shift}}$ is chosen as $d_{\text{shift}}=[1-(\frac{2}{\pi}\arccos{\frac{|k_P|}{k}})^2]\delta_{BL}$.
As shown in \cite{Park3D}, this choice guarantees a progressive and smooth steering of the vehicle along the path. 

\noindent Instrumental for the next results, we also introduce the look-ahead vector computed on the error $\mathbf{e}$ instead of the radially shifted distance $\mathbf{d}$, that is
\begin{equation}
\label{eq:L0}
\mathbf{\hat{L}}_0:=\mathbf{\hat{L}}{|}_{\mathbf{d}=\mathbf{e}}\,.
\end{equation}

\section{THE LOWER WIND CASE}\label{sec:SLOWERWIND}
\noindent In this section, we consider the slower wind case, i.e. $w^\star<v_M^\star$. Here we design $\mathbf{u}_{\text{slow}}$ as in (\ref{eq:full_u}).

\subsection{Previous solutions and their weaknesses}
\begin{figure}[htbp]
\includegraphics[trim={0 2cm 0 2.1cm},clip=true,width=0.5\textwidth]{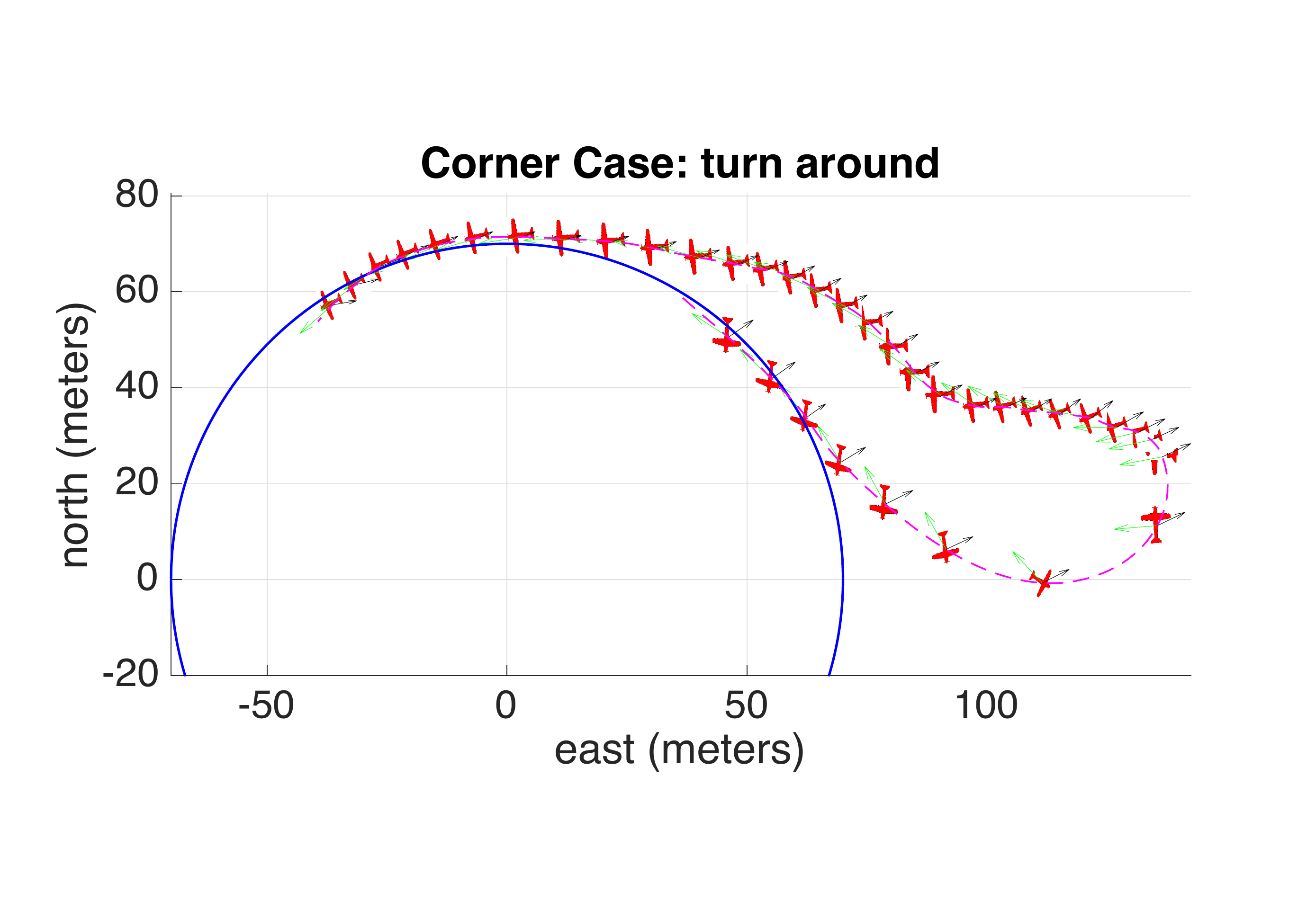}
\caption{Initially, the ground speed is almost aligned with the look-ahead vector, hence the aircraft is not commanded to change its attitude and gets carried away by the wind. The aircraft is forced to perform a complete turn around to get back on track. }
\label{turn_around_sprite}
\end{figure}

\noindent A simple and commonly used approach to achieve path convergence with any wind, similar to that shown in \cite{ParkL1}, is to apply the normal acceleration command ${\mathbf{a}_N^M}=k(\mathbf{v}_G\times \mathbf{\hat{L}}) \times \mathbf{v}_G$ such that  $\mathbf{v}_G$ will eventually be aligned with the look-ahead vector $\mathbf{\hat{L}}$.
Though it should be noted that this acceleration command, defined perpendicular to the ground speed vector, is actually applied to the aircraft body-axis; a notable discrepancy for smaller/slower systems. This approach also presents non-easily predictable behaviours: as an example, it could happen what is shown in Figure \ref{turn_around_sprite}.

  

\subsection{Proposed strategy}\label{slower_winds}
\noindent Here the goal is to find the control input $\mathbf{u}_{\text{slow}}$ that satisfies the requirements in \ref{eq:goals:path-tracking_wind}. We first find a \emph{basic} control input, called $\mathbf{u}_e$, and improve on that to obtain $\mathbf{u}_{\text{slow}}$. To this end, we are going to reason in steady state, i.e.
\begin{equation}
\label{eq:steady_state}
\begin{cases}
\mathbf{e}=0\\
\mathbf{\hat{T}_G}=\mathbf{\hat{T}_P}\\
\frac{d\mathbf{\hat{T}}_G}{dt}=\frac{d\mathbf{\hat{T}}_P}{dt}
\end{cases}
\end{equation}

\subsubsection{Initial control input}
\begin{figure}[htbp]
\centering
\includegraphics[width=0.36\textwidth]{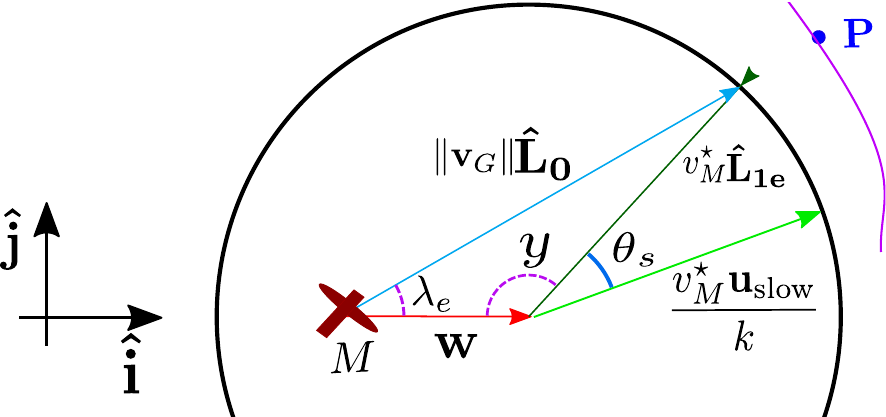}
\caption{$\mathbf{\hat{L}}_0$ is the desired direction for $\mathbf{v}_G$. $\mathbf{\hat{L}}_{1e}$ is the target aircraft orientation. $\mathbf{\hat{u}}_{\text{slow}}$ is the applied control input. $\lambda_e$ and $y$ are defined in (\ref{eq:lambdae_y}) and $\theta_s$ is defined in (\ref{eq:thetas})}
\label{ink_base_slower_winds}
\end{figure}

Here we are going to satisfy the first two requirements in \eqref{eq:goals:path-tracking_wind}.
It is useful to consider the geometry of the problem shown in Figure \ref{ink_base_slower_winds} and introduce the following angles, using basic trigonometric relations

\begin{equation}
\label{eq:lambdae_y}
\begin{cases}
\lambda_e=\arccos{\mathbf{\hat{w}}\cdot \mathbf{\hat{L}}_0}\\ 
y=\arccos{-\mathbf{\hat{w}}\cdot \mathbf{\hat{L}}_{1e}}=\pi-\lambda_e-\arcsin{(\frac{w^\star\sin{(\lambda_e)}}{v_M^\star})}
\end{cases}
\end{equation}
where $\mathbf{\hat{L}}_{1e}$ is an unknown target orientation for the aircraft to be computed. It should be noted that these angles are not defined in case $\mathbf{w}=0$.
We now aim to satisfy the first two requirements stated in \eqref{eq:goals:path-tracking_wind} through the choice of a basic control input
\begin{equation}
\label{eq:ue}
\mathbf{u}_e=k\mathbf{\hat{L}}_{1e}
\end{equation}
\noindent To find such a command, we assumed to already be at the Position/Orientation steady state condition.  Since we assume to be on the path with the desired orientation, then  $k(\mathbf{v}_M \times \mathbf{\hat{L}}_{1e}) \times \mathbf{v}_M=0$, meaning that $\mathbf{\hat{T}}_M=\mathbf{\hat{L}}_{1e}$ ($\mathbf{\hat{T}}_M=-\mathbf{\hat{L}}_{1e}$ would be an unstable equilibrium, as shown in \cite{Park3D}).\\
The natural choice for the desired groundspeed direction is $\mathbf{\hat{L}}_0$, as it was defined in (\ref{eq:L0}).
Note that $\mathbf{\hat{L}}_0{|}_{\mathbf{e}=0}=\mathbf{\hat{T}_P}$.
We need to find the desired direction $\mathbf{\hat{L}}_{1e}$ for the aircraft by solving the geometry shown in Figure \ref{ink_base_slower_winds}, which means solving the following equation in $\mathbf{\hat{L}}_{1e}$:
\begin{equation}
\frac{\mathbf{w}+v_M^\star\mathbf{\hat{L}}_{1e}}{\|\mathbf{w}+v_M^\star\mathbf{\hat{L}}_{1e}\|}=\mathbf{\hat{L}}_0
\end{equation}
The solution, in terms of the angles defined in \eqref{eq:lambdae_y}, is
\begin{equation}
\begin{cases}
\mathbf{\hat{L}}_{1e}=\text{sign}([\mathbf{\hat{w}}\times \mathbf{\hat{L}}_0]\cdot \mathbf{k})\text{rot}(-\mathbf{\hat{w}},-y)\qquad ~w^\star>0\\
\mathbf{\hat{L}}_{1e}=\mathbf{\hat{L}}_0\qquad \qquad \qquad \qquad \qquad \qquad \quad w^\star=0
\end{cases}
\end{equation}
where $\text{rot}(a,\theta)$ is the function that rotates vector $a \in \mathbb{R}^3$ by angle $\theta\in \mathbb{R}$ around the vertical axis $\mathbf{k}$. The basic $\mathbf{u}_e$ will be improved in \ref{subsub:thetas} to obtain curvature convergence.
\subsubsection{Improvement of the control input to satisfy the curvature convergence requirement}
\label{subsub:thetas}
In order to satisfy the curvature convergence requirement, we need to force the correct amount of steady-state centripetal acceleration to the aircraft. This can be done by firstly reasoning on what additional acceleration should be imposed to the $\mathbf{v}_G$ vector (which we call $\|{\mathbf{a}_N^G}_{\text{res}}\|$), then by mapping to the actual aircraft control input.
The function $\|{\mathbf{a}_N^G}_{\text{res}}\|(\cdot)$ should satisfy the steady-state curvature requirement, i.e.:
\begin{equation}
\label{eq:a_res}
\begin{aligned}
&\|{\mathbf{a}_N^G}_{\text{res}}\|{~|}_{\mathbf{e}=0}=\|k_P\|\|\mathbf{v}_G\|^2=\\
&=k\|\mathbf{v}_G\|^2\| (\mathbf{\hat{T}}_P \times \mathbf{\hat{L}_{|\mathbf{d}|=d_{\text{shift}}}}) \times \mathbf{\hat{T}}_P\|
\end{aligned}
\end{equation}
In addition to that, when mapped to the actual control input, it should preserve convergence. Inspired by \cite{Park3D} and equation (\ref{eq:a_res}), we claim that the following function is a suitable choice:
\begin{equation}
\label{eq:a_N^G}
\|{\mathbf{a}_N^G}_{\text{res}}\|=k \|\mathbf{v}_G\|^2 \|(\mathbf{\hat{L}}_0 \times \mathbf{\hat{L}}) \times \mathbf{\hat{L}}_0\|
\end{equation}
with $\mathbf{\hat{L}}$ and $\mathbf{\hat{L}}_0$ as defined in \ref{sub:Nominal Solution}. Indeed, (\ref{eq:a_res}) is satisfied by definition ($\mathbf{\hat{L}}_0|_{\mathbf{e}=0}=\mathbf{\hat{T}}_P,~\mathbf{d}|_{\mathbf{e}=0}=\mathbf{d}_{\text{shift}}$), and convergence appears to be preserved.

\begin{figure}[thpb]
\centering
\includegraphics[width=0.48\textwidth]{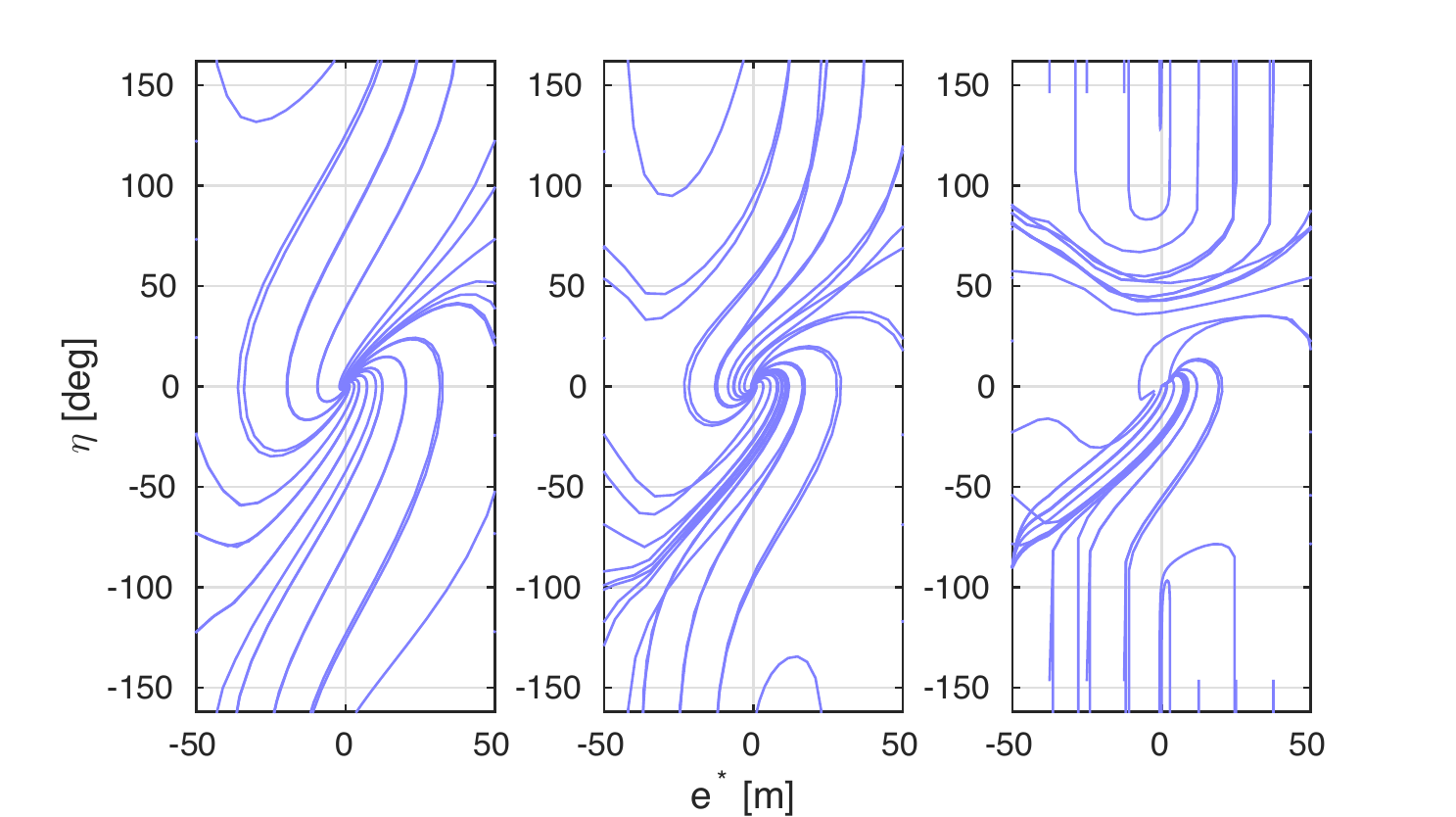}
\caption{Phase portraits of the proposed lower wind solution for $w^\star$=\SI{0}{\meter\per\second} (\textbf{left}), \SI{7}{\meter\per\second} (\textbf{middle}), and \SI{13.5}{\meter\per\second} (\textbf{right}), respectively. The tracking anglular error $\eta=\text{atan2} \left(\mathbf{\hat{T}}_{P_y},\mathbf{\hat{T}}_{P_x}\right)-\text{atan2} \left(\mathbf{\hat{T}}_{G_y},\mathbf{\hat{T}}_{G_x}\right)\in\left[-\pi,\pi\right]$ is compared with the signed, one-dimensional cross-track error $e^*=\mathbf{e}\cdot \frac{\mathbf{r}_M}{\|\mathbf{r}_M\|}$ to show algorithm convergence within the bounds of $\delta_{BL}=50\text{ m}$, for $k=0.05$, $R=100\text{ m}$, and $v_M^\star$=\SI{14}{\meter\per\second}.}
\label{fig:phase-portrait}
\end{figure}

\begin{figure}[thpb]
\includegraphics[trim={0 2cm 1cm 1.9cm},clip=true,width=0.45\textwidth]{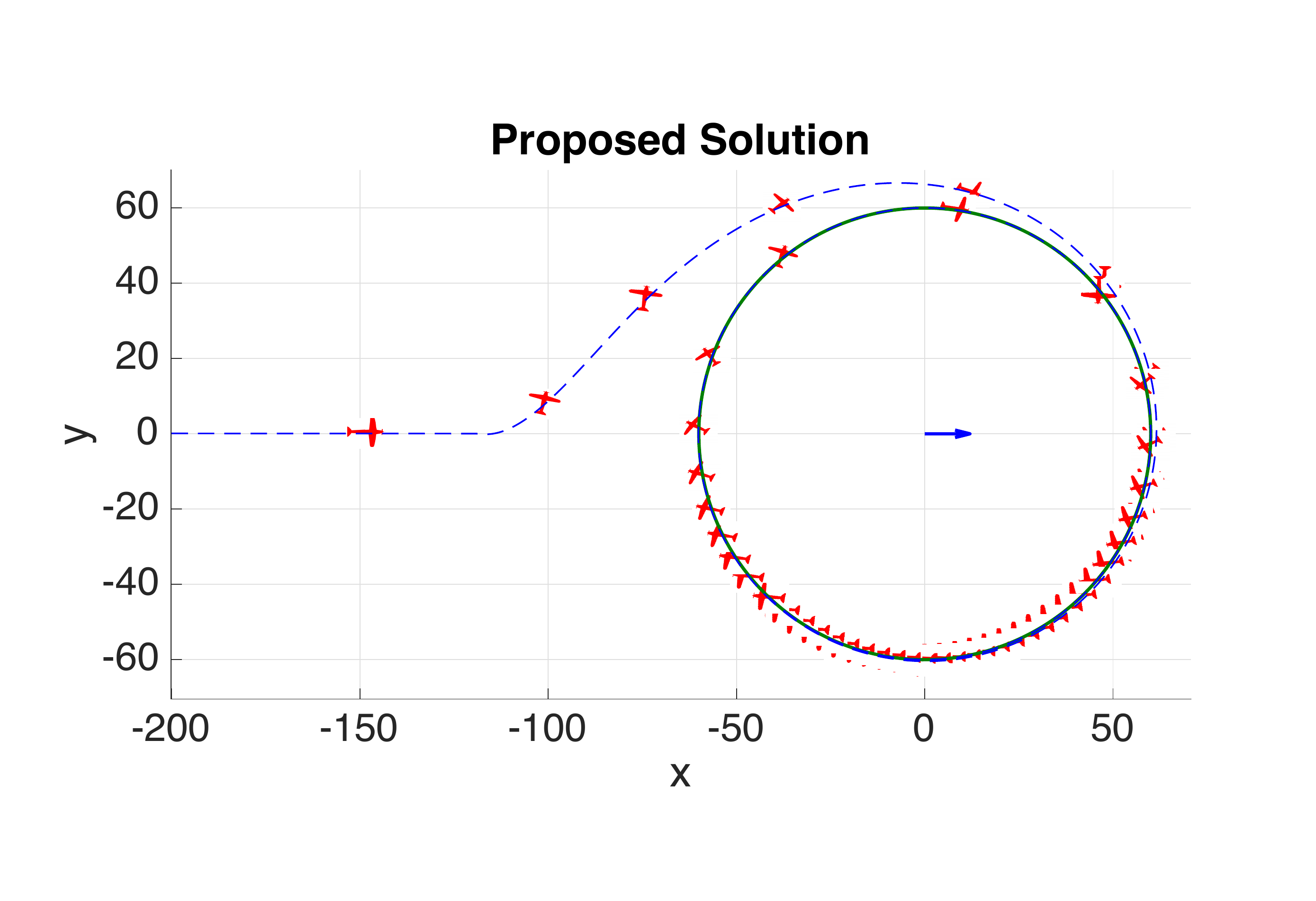}
\caption{Airspeed 14 m/s. Windspeed 12 m/s. The proposed solution lets the vehicle achieve the goals in (\ref{eq:goals:path-tracking_wind}).}
\label{Proposed_slow_winds}
\end{figure}

\textbf{Mapping to the control input.} We show that additional centripetal acceleration for the aircraft can be achieved by rotation of the basic control input $\mathbf{u}_e$ (\ref{eq:ue}) through a properly shaped angle function $\theta_s(\cdot)$. Notice indeed that, for $\theta_s^\star \in \mathbb{R}$:
\begin{equation}
\label{eq:theta_effect}
[(\mathbf{v}_M\times \text{rot}(\mathbf{u}_e,\theta_s^\star))\times \mathbf{v}_M] |_{\text{angle}(\mathbf{v}_M,\mathbf{u}_e)=0}=k {v_M^\star}^2\sin(\theta_s^\star)
\end{equation}
Where we assumed the vehicle direction to coincide with $\mathbf{u}_e$. Since $\|{\mathbf{a}^G_N}_{\text{res}}\|$ is applied to the ground-speed vector $\mathbf{v}_G$, and remembering that for any normal acceleration it holds $\mathbf{a}_N=\vec{\Omega} \times \mathbf{V}$, where $\vec{\Omega}$ is the angular speed vector and $\mathbf{V}$ is the linear speed vector, then it holds through derivation w.r.t. time of $\lambda_e$ defined in (\ref{eq:lambdae_y}) (which indicates the $\mathbf{v}_G$ orientation):
\begin{equation}
\label{eq:lambdae_dot}
\dot{\lambda}_e=\frac{\|{\mathbf{a}_N^G}_{\text{res}}\|}{\|\mathbf{v}_G\|}\text{sign}(\kappa_P)\\
\end{equation}
Still assuming that $\text{angle}(\mathbf{v}_M,\mathbf{u}_e)=0$, noticing that angle $y$ defined in (\ref{eq:lambdae_y}) indicates the aircraft body-axis to which we apply the control input, and considering equation (\ref{eq:theta_effect}), it holds: 
\begin{equation}
\label{eq:approx_doty}
\dot{y}=\frac{d}{dt}\left(\text{angle}(\mathbf{v}_M,\mathbf{w})|_{\text{angle}(\mathbf{v}_M,\mathbf{u}_e)=0}\right)=k {v_M^\star}\sin(\theta_s)
\end{equation}
Also, deriving the second equation of (\ref{eq:lambdae_y}) w.r.t time, it holds:
\begin{equation}
\label{eq:phys_doty}
\begin{aligned}
\dot{y}&=-\dot{\lambda}_e-\frac{w^\star\cos{(\lambda_e)}\dot{\lambda}_e}{v_M^\star\sqrt{1-(\frac{w^\star\sin(\lambda_e)}{v_M^\star})^2}}\\
\end{aligned}
\end{equation}
Hence plugging (\ref{eq:a_N^G}) into (\ref{eq:lambdae_dot}) and  (\ref{eq:lambdae_dot}) into (\ref{eq:phys_doty}), we can compare equations (\ref{eq:approx_doty}) and (\ref{eq:phys_doty}) to obtain:

\begin{equation}
\label{eq:thetas}
\scalebox{0.87}{
$\frac{\theta_s}{\text{sign($\kappa_P$)}}=\arcsin{\left[\text{sat}\left(-\frac{\|\mathbf{v}_G\|\|(\mathbf{\hat{L}}_0 \times \mathbf{\hat{L}})\times \mathbf{\hat{L}_0}\|}{v_M^\star}\left(1+\frac{w^\star\cos{(\lambda_e)}}{\sqrt{{v_M^\star}^2-\left(w^\star\sin(\lambda_e)\right)^2}}\right)\right)\right]}$}
\end{equation}
\normalsize
Where the saturation function bounds the argument between -1 and 1: this is needed because of the assumption $\text{angle}(\mathbf{v}_M,\mathbf{u}_e)=0$, i.e. during the transient we might ask for residual accelerations that are higher than in steady-state. This doesn't ruin convergence, as the vehicle will keep turning until the $\|\mathbf{v}_G\|$ eventually decreases and $\theta_s$ can smoothly steer the trajectory curvature to the path curvature. In the end we apply:
\begin{equation}
\label{eq:ufromue}
\mathbf{u}_{\text{slow}}=\text{rot}(\mathbf{u}_e,\theta_s)
\end{equation}
 \noindent $\mathbf{u}_e$ defined as in (\ref{eq:ue}), $\theta_s$ as in (\ref{eq:thetas}), 
  so that the goals in \eqref{eq:goals:path-tracking_wind} are satisfied. We report in Figure  \ref{fig:phase-portrait} a phase portrait showing global convergence in numerical simulations for a large variety of initial conditions and different windspeeds. That said, attractiveness to the equilibrium is not formally proved in this paper.\\
\noindent In Figure \ref{Proposed_slow_winds}, we can observe the performance of the algorithm for strong constant wind, still slower than the airspeed.
\textbf{Choice of $\mathbf{k}$}. In order for the algorithm to keep null error in steady state, we  have the lower bound:
\begin{equation}
\label{low_bound}
k> \max_{|k_P|}\left(1+\frac{w^\star}{v_M^\star}\right)^2{|k_P|}
\end{equation}
Similarly to \cite{Park3D}, the derivation considers the highest acceleration we need in the worst case scenario ($\mathbf{\hat{L}}_0=\mathbf{\mathbf{\hat{w}}}$, $\mathbf{v}_M\parallel \mathbf{w}$).
\section{THE HIGHER WIND CASE}
\label{sec:higher-winds}
\begin{figure}[thpb]
\includegraphics[width=0.43\textwidth]{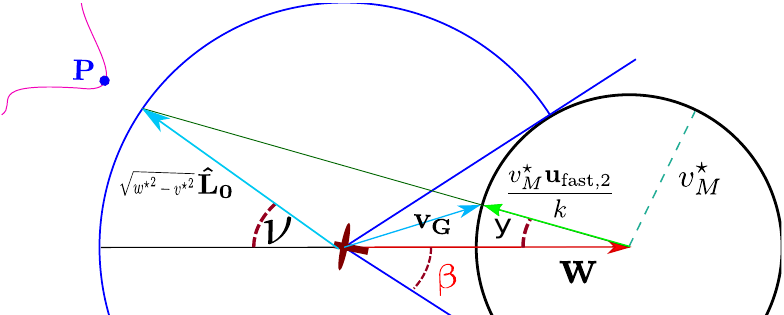}
\caption{$w^\star>v_M^\star$ case. $\nu$ is the angle between $-\mathbf{w}$ and the desired groundspeed $\mathbf{\hat{L}}_0$, $y$ as in Figure 3, $\mathbf{v}_G$ is the actual groundspeed that we achieve, $\mathbf{u}_{\text{fast,2}}$ is the chosen control input.}
\label{ink_base_higher_winds}
\end{figure}

In this section we design $\mathbf{u}_{\text{fast,1}}$ and $\mathbf{u}_{\text{fast,2}}$ introduced in (\ref{eq:full_u}). Let us define the desired direction for the groundspeed $\mathbf{\hat{L}}_0$ as in equation \eqref{eq:L0}, and the corresponding basic control input $\mathbf{u}_e$ as in equation \eqref{eq:ue}. It is convenient to reason considering the angles introduced in \eqref{eq:lambdae_y}: refer to Figure \ref{ink_base_higher_winds} for a better visualization.


\subsection{Solution for $\mathbf{\hat{L}}_0$ feasible, i.e. $\lambda \leq \beta$}
\label{subsec:INSIDECONE}
As the desired groundspeed direction $\mathbf{\hat{L}}_0$ is feasible, we reason as in \ref{slower_winds}: choose the basic control input $\mathbf{u}_e$ as in \eqref{eq:ue} and rotate it by a proper angle in order to achieve curvature convergence: this would mean $\mathbf{u}_{\text{fast,1}}=\mathbf{u}_{\text{slow}}$, and doing so we would achieve curvature convergence as long as the $\mathbf{\hat{L}}_0$ is still feasible. However, with usual shapes for the target curved path, at some point the desired direction \emph{will} become infeasible: when this happens, we need the control input not to change abruptly, i.e. to be a continuous function of the desired $\mathbf{\hat{L}}_0$. Since we cannot in general achieve the goals in \eqref{eq:goals:path-tracking_wind}, we  make a slightly different euristic choice for $\mathbf{u}_{\text{fast,1}}$ that guarantees continuity of the commands (as better explained in \cite{arxiv}), while preserving curvature convergence to a good extent as long as the $\mathbf{\hat{L}}_0$ is feasible:
\begin{equation}
\label{eq:thetas2}
\theta_{s2}=\frac{\sqrt{\left[1-(\frac{w^\star\sin{(\lambda_e)}}{v_M^\star})^2\right]}}{\cos{(\lambda_e)}}\theta_s
\end{equation}

\noindent Notice indeed that at the infeasibility boundary
$\theta_{s2}|_{\lambda_e=\arcsin{v_M^\star {w^\star}^{-1}}}=0$ and at the slower wind case boundary $\theta_{s2}|_{w^\star=v_M^\star}=\theta_s$, $\theta_s$ as in (\ref{eq:thetas}). So, in the end,
\begin{equation}
\label{eq:ufast1}
\mathbf{u}_{\text{fast,1}}=\text{rot}(\mathbf{u}_e,\theta_{s2})
\end{equation}

\subsection{Strategy for $\mathbf{\hat{L}}_0$ infeasible, i.e. $\lambda > \beta$}
\label{subsec:OUTSIDECONE}
We define an \emph{infeasibility paramater} $\alpha_{\text{out}}$ and a \emph{safety function} $\sigma_{\text{safe}}(\alpha_{\text{out}})$ as follows:
\begin{equation}
\alpha_{\text{out}}=\frac{\lambda-\beta}{\pi-\beta} \qquad  \sigma_{\text{safe}}=\frac{\frac{\pi}{2}-\beta-y(\alpha_{\text{out}})}{\frac{\pi}{2}-\beta}
\end{equation}
both indices have maximum value equal to 1. When $\sigma_{\text{safe}}=1$, it means that we act conservatively and choose $\frac{\mathbf{u}_{\text{fast,2}}}{k}=-\mathbf{\hat{w}}$: this has to happen only in the absolutely worst scenario of $\mathbf{\hat{L}}_0=-\mathbf{\hat{w}}$, which corresponds to the maximum $\alpha_{\text{out}}=1$.  In all the intermediate cases, we want to guarantee a tradeoff between conservatism and tracking performance, i.e we want $\sigma_{\text{safe}}(\alpha_{\text{out}})$ to be increasing with $\alpha_{\text{out}}$.

 
\noindent This can be achieved by finding a proper mapping $f$ from angle $\nu=\pi-\lambda_e$ to angle $y$ in the following form

\begin{equation}
f: \nu \in \left[0,\pi-\beta \right]\rightarrow y \in \left[0,\frac{\pi}{2}-\beta \right]
\end{equation}
\noindent This mapping should  satisfy, at least, these 3 properties:

\begin{equation}
\label{eq:yrequirements}
\begin{aligned}
&f(0)=0\\
&f(\pi-\beta)=\frac{\pi}{2}-\beta\\
&f(a)< f(b)~\forall a>b,~a,b\in \left[0,\frac{\pi}{2}-\beta \right]\\
\end{aligned}
\end{equation}

\begin{figure}[thpb]
      \centering
      \includegraphics[width=0.45\textwidth]{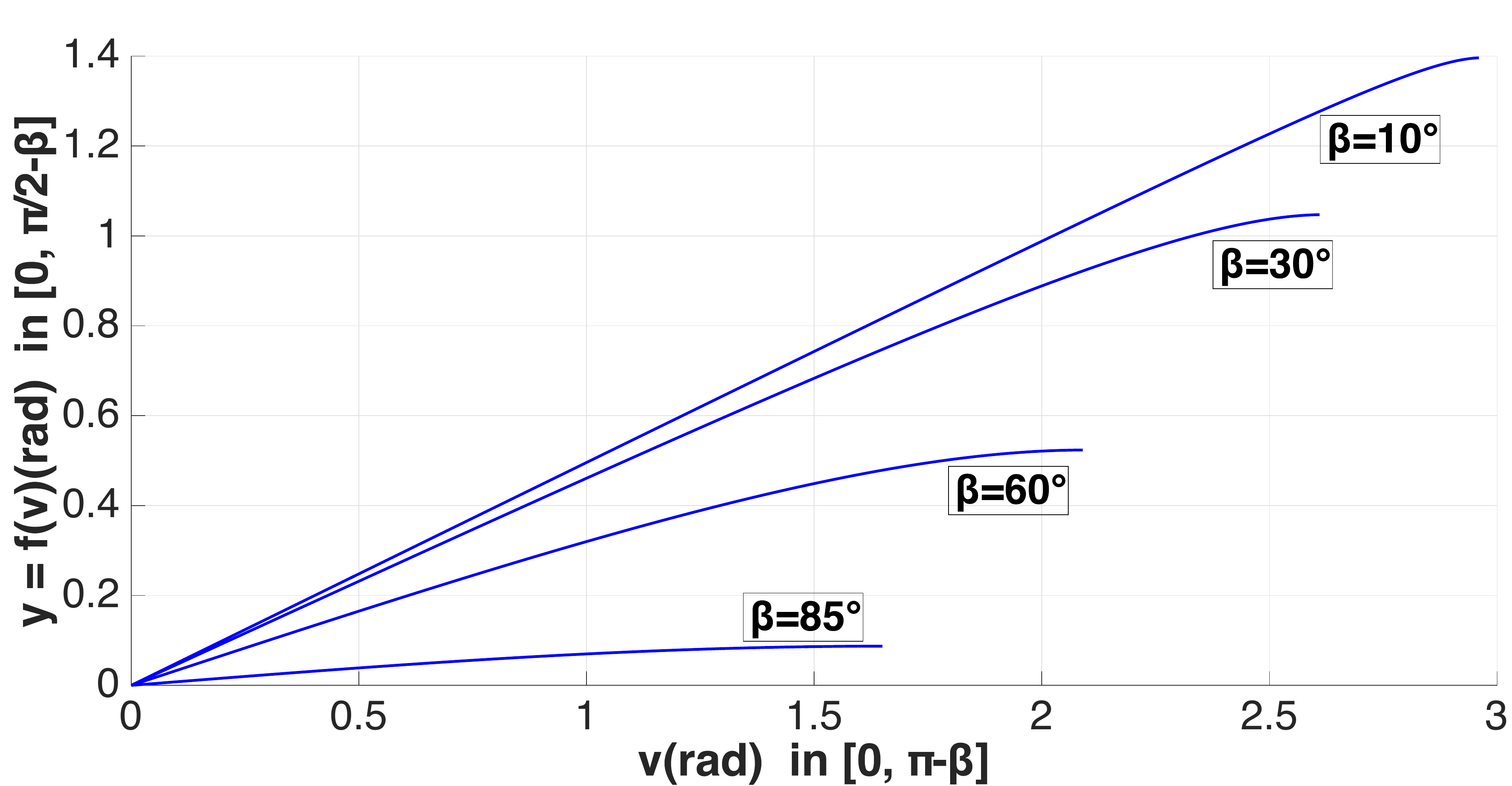}
      \caption{Proposed mapping $y=f(\nu)$ for different values of $\beta$}
\label{fig:proposed_mapping}
\end{figure}

\noindent The first requirement is to guarantee that $\sigma_{\text{safe}}=1$ when $\mathbf{\hat{L}}_0=-\mathbf{w}$. The second one is a boundary condition to guarantee that the input is continuous to the $\mathbf{\hat{L}}_0$ switching from being feasible to infeasible (or vice versa).
The third requirement is for finding a tradeoff between safety and performance: put in words, the more the $\mathbf{\hat{L}}_0$ is infeasible for the groundspeed, the more we want to turn against the wind and wait for it to stop.\\
%
By looking at Figure \ref{ink_base_higher_winds}, a natural choice that follows geometric intuition and is coherent with the requirements that we have just stated, is
\begin{equation}
\label{eq:ufast2}
\mathbf{u}_{\text{fast,2}}=k \frac{\sqrt{{w^\star}^2-{v_M^\star}^2}\mathbf{\hat{L}}_0-\mathbf{w}}{\|\sqrt{{w^\star}^2-{v_M^\star}^2}\mathbf{\hat{L}}_0-\mathbf{w}\|}
\end{equation}
In terms of the mapping that has been defined before, this choice corresponds to
\begin{equation}
\label{eq:higher-wind-y}
f(\nu=\pi-\lambda)=y=\arcsin{\frac{\sin{\nu}\cos{\beta}}{\sqrt{1+\cos^2{\beta}+2\cos{\beta}\cos{\nu}}}}
\end{equation}
this mapping satisfies the requirements \eqref{eq:yrequirements}, as can be easily verified by substition and derivation with respect to $\nu$. For clarity, the function is plotted in Figure \ref{fig:proposed_mapping} for different values of the wind-cone opening angle $\beta$

\begin{figure}[thpb]
\includegraphics[width=0.45\textwidth]{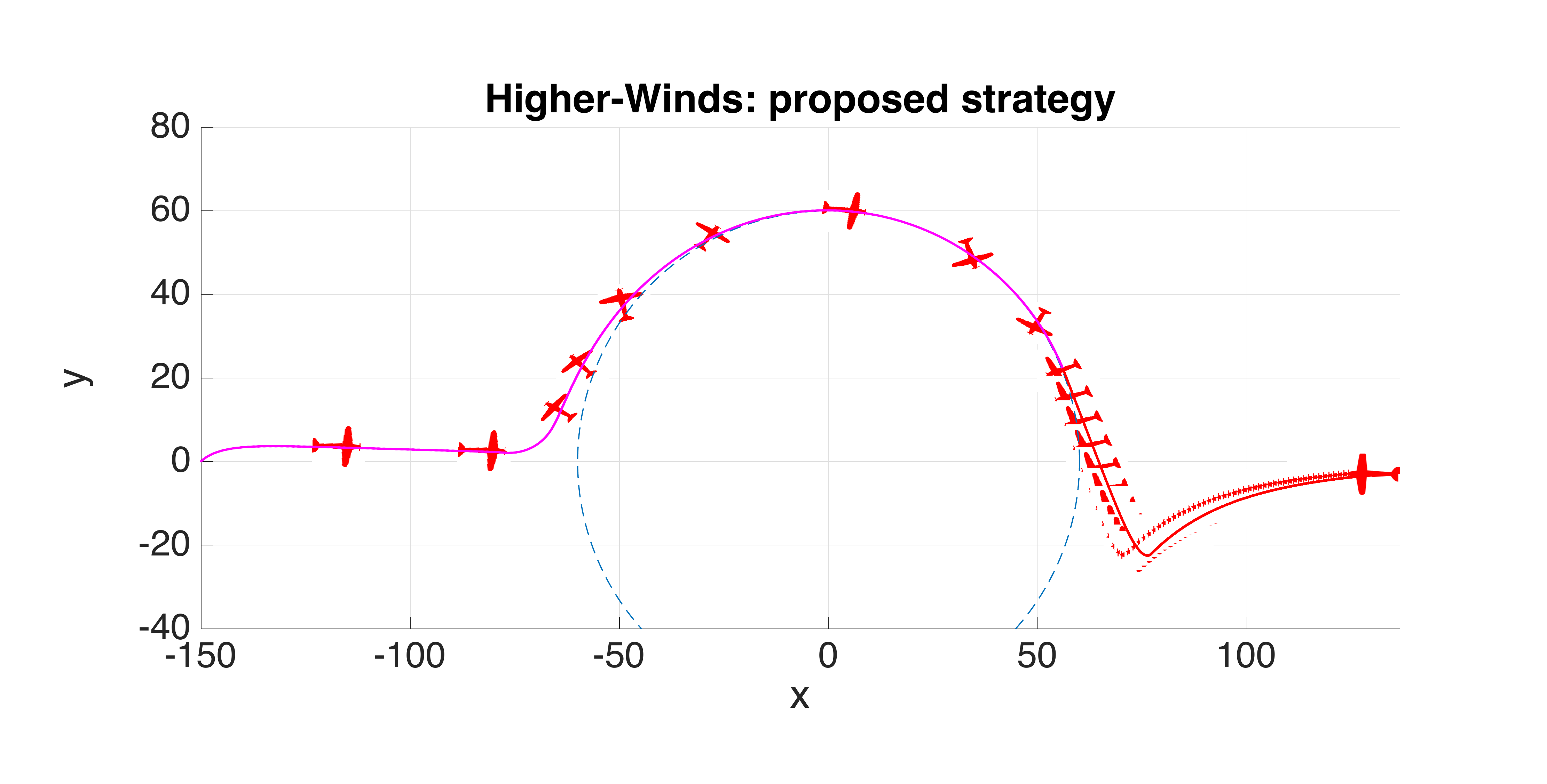}
\caption{Windspeed is $16~m/s$, airspeed is $14~m/s$. The proposed control input is used. Magenta line: feasible desired direction. Red line: infeasible desired direction.}
\label{circle_aircraftSprite_21ms}
\end{figure}

\noindent In Figure \ref{circle_aircraftSprite_21ms} the performance of the algorithm is shown.

\begin{figure}[thpb]
      \centering
      \includegraphics[width=0.46\textwidth]{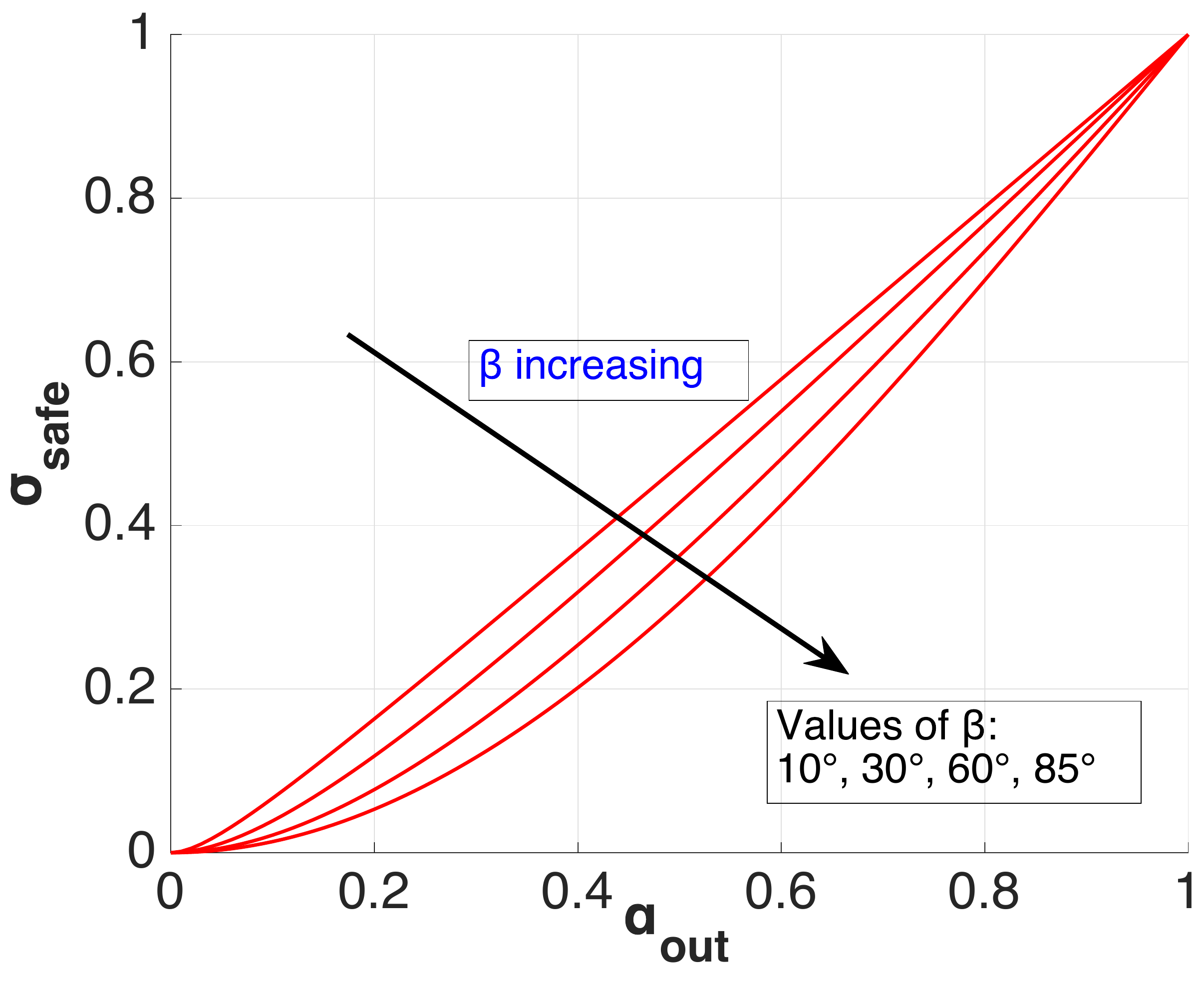}
      \caption{Infeasibility/safety relationship, for different $\beta$s}
\label{fig:performance}
\end{figure}

It is also worth highlighting the tradeoff introduced between performance and safety (\textbf{incremental safety}) by computing the safety function $\sigma_{\text{safe}}(\alpha_{\text{out}})$ :
\begin{equation}
\sigma_{\text{safe}}=\frac{\beta-\frac{\pi}{2}+\arcsin{\left(\frac{\sin{(\pi-\beta-(\pi-\beta)\alpha_{\text{out}})}\cos{\beta}}{\sqrt{1+\cos{\beta}^2+2\cos{\beta}\cos{(\pi-\beta-(\pi-\beta)\alpha_{\text{out}})}}}\right)}}{\beta-\frac{\pi}{2}}
\end{equation}
 as is also shown in Figure \ref{fig:performance}.

\preprintswitch{A proof of convergence properties is made available in \cite{arxiv}}. 
\section{CONTINUITY}
\label{sec:CONTINUITY}
In realistic scenarios, the wind is not going to be constant, but will likely switch between $w^\star<v_M^\star$ and $w^\star>v_M^\star$ several times. Not only that, the path is going to be curved, so the desired direction for the groundspeed $\mathbf{\hat{L}}_0$ is going to switch between being feasible and infeasible. All these switchings mean that it is very important for the command input $\mathbf{u}$ to be continuous to changing winds and changing $\mathbf{\hat{L}}_0$.\\
The control input was derived separately for the three subcases (slower winds, higher winds with feasible desired direction, higher winds with infeasible desired direction) in  \ref{sec:SLOWERWIND}, \ref{subsec:INSIDECONE}, \ref{subsec:OUTSIDECONE}. \preprintswitch{Continuity properties are discussed in more detail in the accompanying preprint in \cite{arxiv}.}{We want to show here that the complete control input

\begin{equation}
\mathbf{u}=
\begin{cases}
\mathbf{u}_{\text{slow}},\qquad w^\star \leq v_M^\star\\
\mathbf{u}_{\text{fast,1}}, \qquad w^\star > v_M^\star , ~\lambda\leq\beta\\
\mathbf{u}_{\text{fast,2}}, \qquad w^\star > v_M^\star, ~\lambda>\beta
\end{cases}
\end{equation}
indeed guarantees continuity in this sense.\\
 
\begin{itemize}
\item \textbf{Switching between $\mathbf{u}_{\text{slow}}$ and $\mathbf{u}_{\text{fast,1}}$}: this happens as the wind passes from $w^\star<v_M^\star$ to $w^\star>v_M^\star$. Let $t^*$ be the boundary time instant in which $w^\star(t^*)=v_M^\star(t*)$. Also, in this case, $\lambda_e(t^*)\leq\frac{\pi}{2}$. Looking at the formulation for $\theta_s$ and $\theta_{s2}$ in \eqref{eq:thetas2}, we have:
\begin{equation}
\theta_{s2}{|}_{w^\star=v_M^\star}=\theta_{s}{|}_{w^\star=v_M^\star}
\end{equation}
and so $\mathbf{u}_{\text{slow}}(t^*)=\mathbf{u}_{\text{fast,1}}(t^*)$

so the command $\mathbf{u}(t)$ is continuous at this boundary condition

\item \textbf{Switching between $\mathbf{u}_{\text{slow}}$ and $\mathbf{u}_{\text{fast,2}}$}: this happens as the wind passes from $w^\star<v_M^\star$ to $w^\star>v_M^\star$. Let $t^*$ be the boundary time instant in which $w^\star(t^*)=v_M^\star(t*)$. Also, in this case, $\lambda_e(t^*)>\frac{\pi}{2}$.
 Solving the geometry in \ref{ink_base_slower_winds}, we have that $\mathbf{u}_e=-\mathbf{w}$. Since $\beta(t^*)=\frac{\pi}{2}$, this implies that $y(t^*)=0$ as computed in \eqref{eq:higher-wind-y}: so $\mathbf{u}_{\text{fast,2}}(t^*)=-\mathbf{w}$ as well.  Assuming $\mathbf{\hat{T}}_M\approx \mathbf{\hat{L}}_{1e}$, which is the case after some transient, we have that $\|\mathbf{v}_G\|\approx 0$. So by \eqref{eq:dotlambday}, we have $\theta_s \approx 0$, implying 
\begin{equation}
\mathbf{u}_{\text{slow}}(t^*)\approx -\mathbf{w} \approx \mathbf{u}_{\text{fast,2}}(t^*)
\end{equation}


\item \textbf{Switching between $\mathbf{u}_{\text{fast,1}}$ and $\mathbf{u}_{\text{fast,2}}$}: this happens when $w^\star>\mathbf{v}_M$ and $\mathbf{\hat{L}}_0$ passes from being feasible to being infeasible. Let $t^*$ be the boundary time instant. Then
\begin{equation}
w^\star(t^*)\sin{\lambda_e(t^*)}=v_M^\star(t^*)
\end{equation}

so $\theta_{s2}(t^*)=0$. This guarantees continuity, as no shifting angle is applied in the infeasible case.

%
%
%
%
\end{itemize}
} 

\begin{figure}[thpb]
      \centering
 \includegraphics[width=0.49\textwidth]{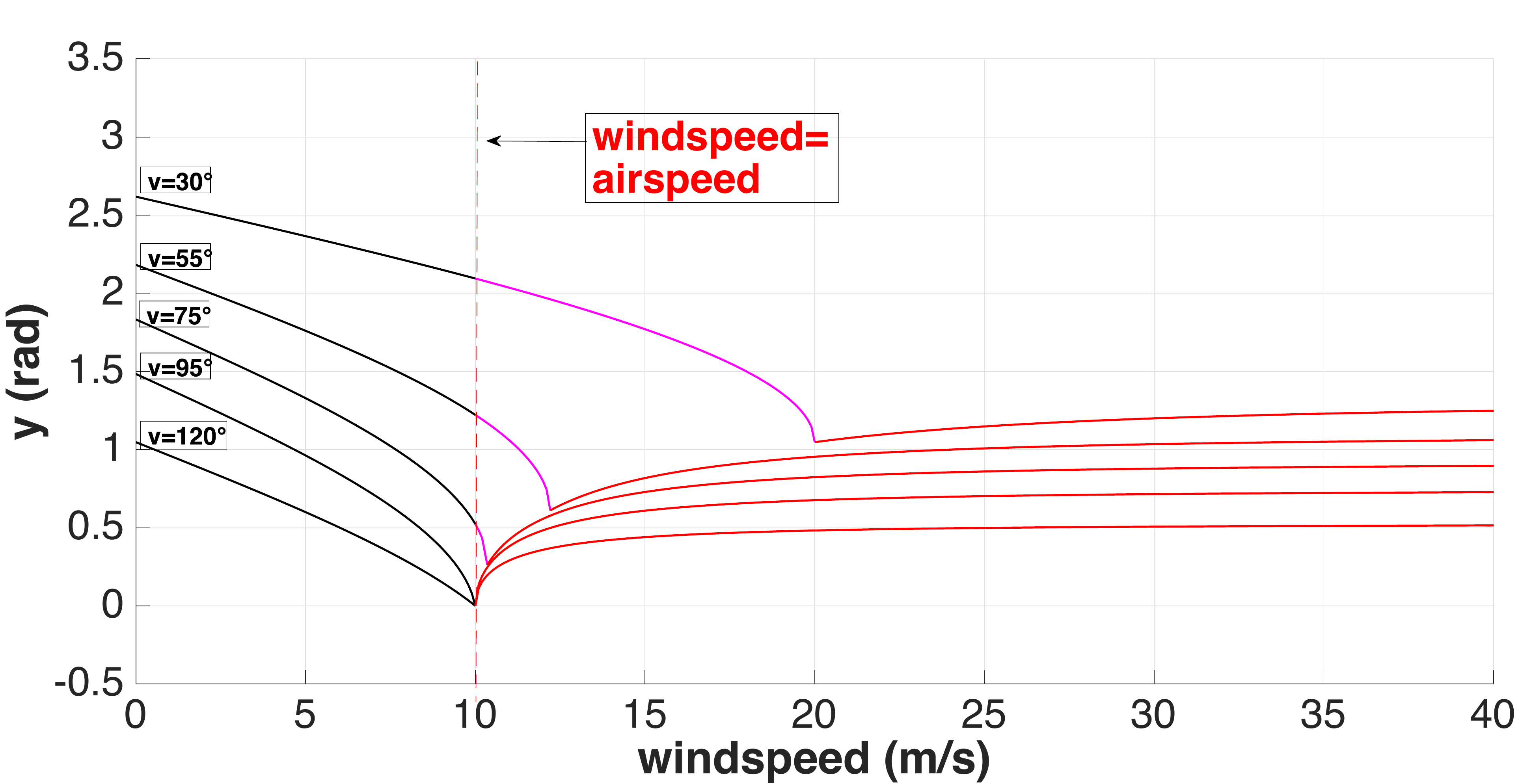}
      \caption{Continuity at $w^\star=v_M^\star$. Black line: slower wind. Magenta: inside the cone. Red: outside the cone.}
\label{fig:wind_varying2}
\end{figure}

In Figure \ref{fig:wind_varying2} we report a plot that highlights the continuity of the input as the wind increases and for a fixed $\mathbf{\hat{L}}_0$. We plot angle $y$ associated with the direction of the control input $\mathbf{u}_{\text{fast,2}}$, for different values of $\nu=\text{angle}(-\mathbf{w},\mathbf{\hat{L}}_0)$.

\subsection{Sinusoidal Winds }
\begin{figure}[thpb]
      \centering
      \includegraphics[trim={10cm 0 10cm 0},clip=true,width=0.4\textwidth]{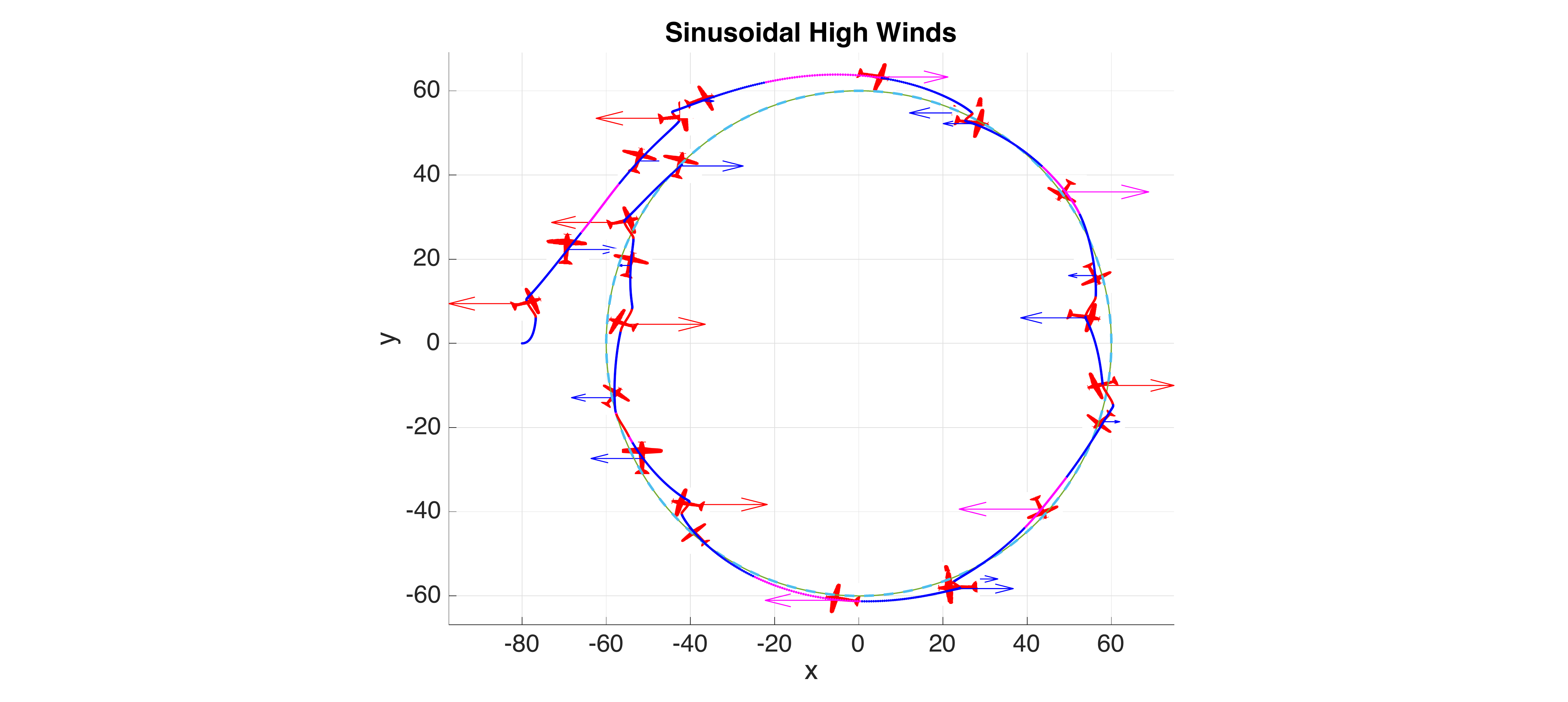}
      \caption{Sinusoidal winds. Blue: $\mathbf{u}_{\text{slow}}$ is applied. Magenta: $\mathbf{u}_{\text{fast,1}}$ is applied. Red: $\mathbf{u}_{\text{fast,2}}$ is applied}
\label{sinusoidal_high}
\end{figure}

As an example of a more realistic varying wind profile, in order to show that the commands do not switch abruptly and are continuous, we consider the case of the wind having this sinusoidal profile 
\begin{equation}
\mathbf{w}(t)=W\sin{(\Omega t)}\begin{bmatrix}
1&0&0
\end{bmatrix}^T
\end{equation}
for some wind pulsation $\Omega$ and amplitude $W>v_M^\star$. The result is shown in Figure \ref{sinusoidal_high}, and the same is shown (more clearly) in an accompanying video\footnote{Sinusoidal wind simulation: <https://www.youtube.com/watch?v=fpV5KkmrrUc>}.

Switching between any couple of the three parts of the control input can happen in this case. The least smooth behaviour, as we have only approximate continuity, is when the switching is between $\mathbf{u}_{\text{slow}}$ and $\mathbf{u}_{\text{fast,2}}$. 



  
\section{FLIGHT RESULTS}
\label{sec:REALFLIGHT}
\begin{figure}[thb] 
      \centering
      \begin{subfigure}[b]{0.24\textwidth}
      	\includegraphics[width=1\textwidth]{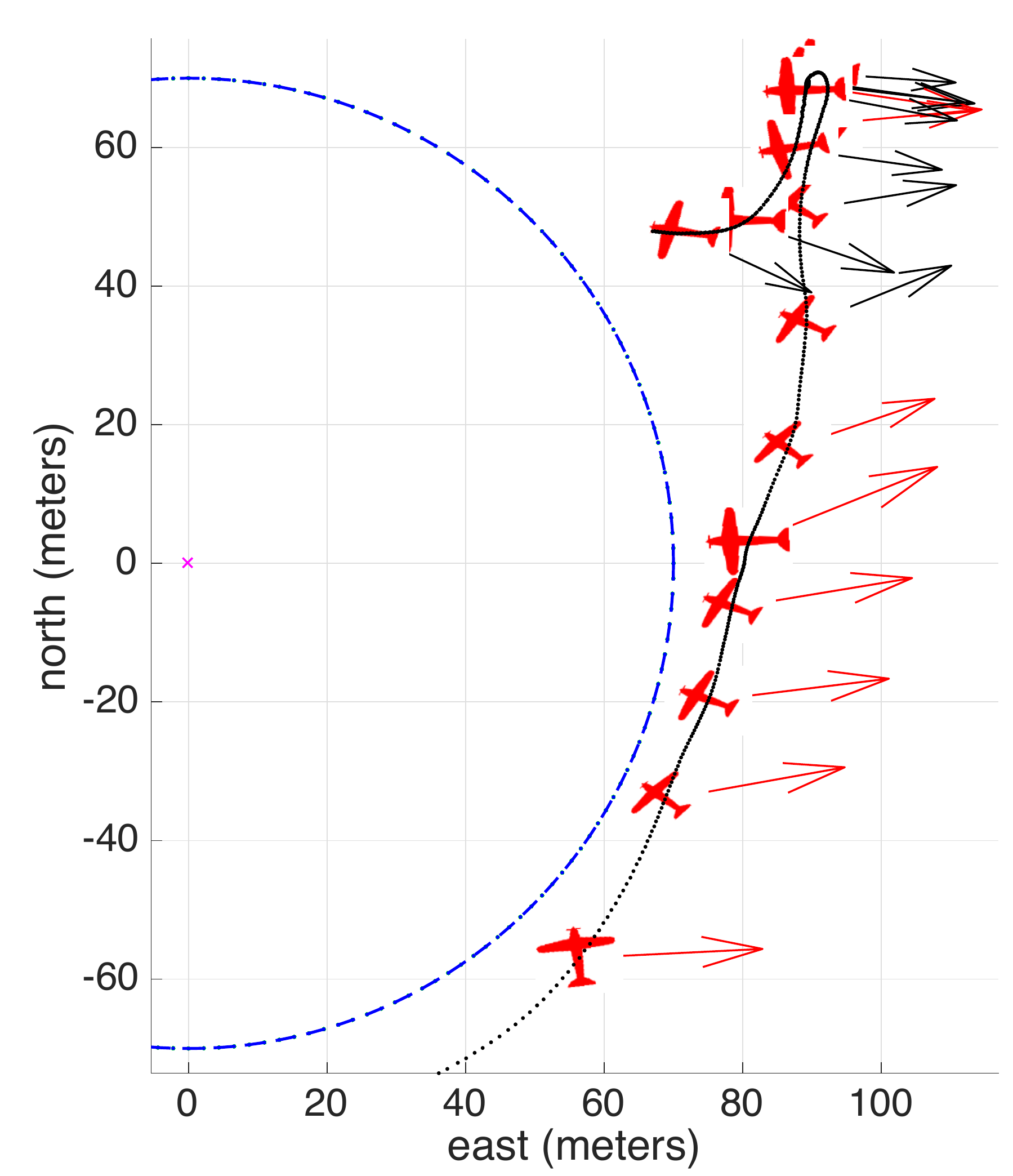}
      	\caption{}
        \label{fig:f01}
    	  \end{subfigure}\begin{subfigure}[b]{0.24\textwidth}
        \includegraphics[width=1\textwidth]{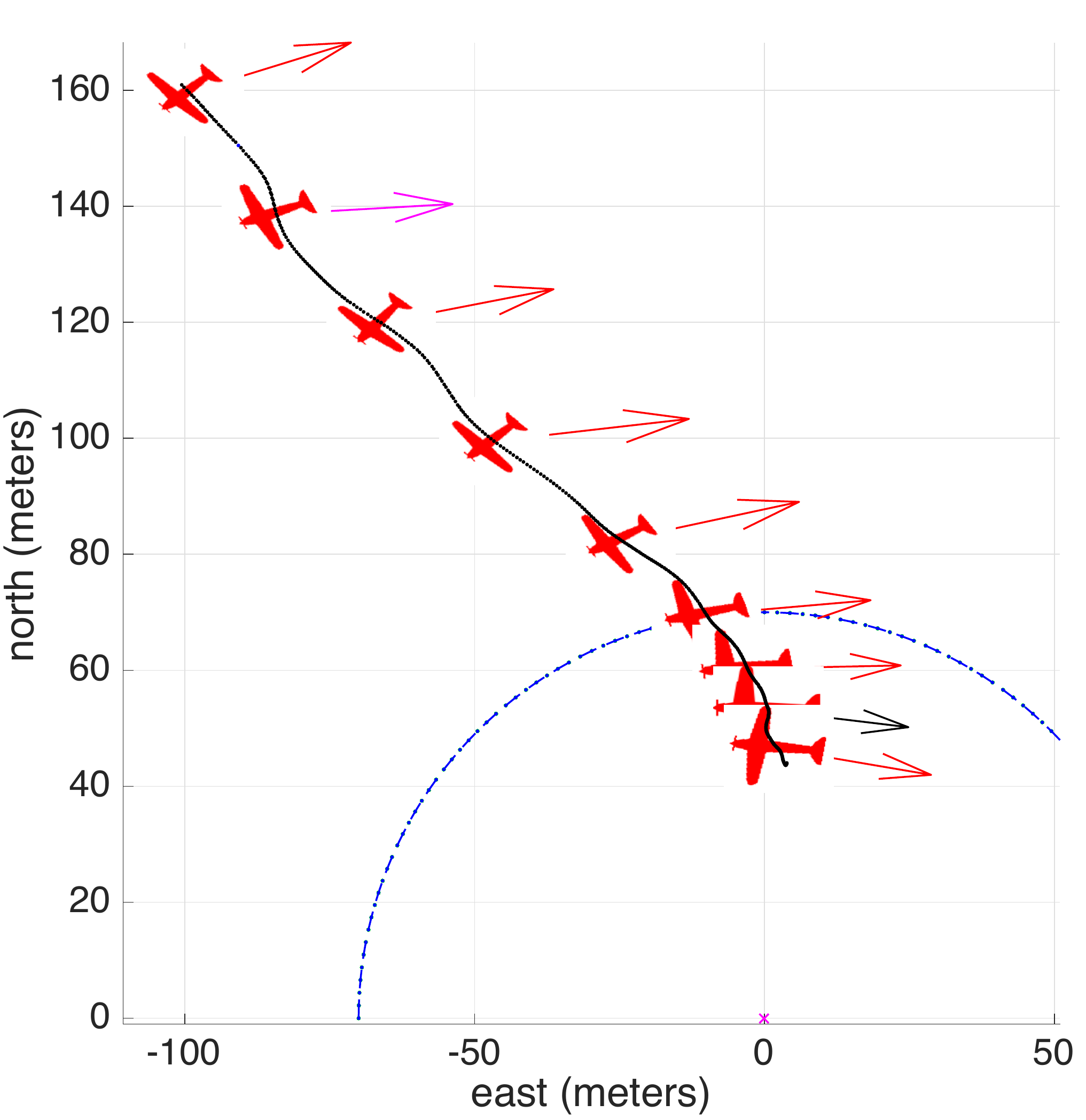}
        \caption{}
        \label{fig:f02}
      \end{subfigure}
      \caption{Windy flight experiments. Red arrow: $w^\star>v_M^\star$}
\label{fig:flight_tests}
\end{figure}

The proposed algorithm was implemented on a Pixhawk Autopilot in C++ and subsequently tested on a small fixed-wing UAV in high wind conditions. In Figure~\ref{fig:flight_tests}, we show the results from the flight tests. The aircraft was commanded to follow a circular trajectory in counter-clockwise direction at a nominal airspeed of \SI{8}{\meter\per\second}. The wind vector is represented in the figures using the following arrow, color scheme: $w^\star<v_M^\star$ (black), $w^\star>v_M^\star\cap(\mathbf{\hat{L}}_0\text{ feasible})$ (magenta), $w^\star>v_M^\star\cap(\mathbf{\hat{L}}_0\text{ infeasible})$ (red). In Figure~\ref{fig:f01}, the UAV can be seen to attempt curvature following despite the infeasible look-ahead direction until a point where the wind speed reduces and allows the start to convergence back to the path. Figure~\ref{fig:f02} shows a wind-stabilized approach towards the trajectory until the point where simply pointing into the wind is the only option to reduce ``runaway" from the track, recall the tracking direction is counter-clockwise.

\section{CONCLUSIONS}
\label{CONCLUSIONS}
In this work, we extended a nonlinear guidance method based on a look-ahead vector, particularly suitable for fixed-wing UAVs, so as to actively take into account the measurements of external flowfields and drastically improve the tracking performance of the vehicle. Taking inspiration from issues that often arise when using small-sized UAVS, such as the maximum achievable airspeed being lower than the windspeed and the commands to the aircraft being discontinuous when treating the higher-wind case as a corner faulty case, the proposed technique considers arbitrarily strong flowfields and is continuous to wind changes.

The slower wind case allows for exact convergence to the path, as all the directions for the ground speed are feasible. The higher-wind case was considered in two separate subcases, by defining the notion of \emph{feasible} and \emph{infeasible} desired ground speed directions for the aircraft. Exact tracking performance in the feasible case was shown preserved, while safety in the infeasible case was demonstrated and bounded to a minimum ``run away" configuration; i.e. we define the concept of \emph{asymptotic safety} for finite paths.

Future work will need to extend the proposed geometric approach to the more general case of 3D paths with 3D winds. A mathematical proof for the convergence with slower winds will also have to be provided.




\preprintswitch{}{
\section*{APPENDIX: Stability Proof for High Wind Case}

We are in the scenario of $w^\star>v_M^\star$.
For finite-length paths, we want to show that we achieve the requirements in (\ref{eq:goals_finitelength}).

\noindent Here we will also consider briefly the case of infinite paths: the only realistic case in UAV application is that of infinite linear paths. 
In this case we want to show that:
\begin{equation}
\label{eq:goals_linear}
\begin{cases}
\underset{t\rightarrow \infty}{\lim}{\mathbf{{a}}_N^M}(t)= 0\\
\underset{t \rightarrow \infty}{\mathbf{\hat{T}}_M} = -\text{rot}(-\mathbf{\hat{w}},f(\frac{\pi}{2}-\mu))
\end{cases}
\end{equation}
\noindent where $\mu=\arccos{\mathbf{\hat{w}}\cdot \hat{\Lambda}}$, $\hat{\Lambda}$ is the direction of the target linear path, mapping  $f: \nu \rightarrow y$ has to be chosen. 
The second requirement in  (\ref{eq:goals_linear}) asks for a trade-off between the linear-path direction and the anti-wind direction for the $\mathbf{\hat{T}}_M$, that results in an efficient direction for the actual $\mathbf{\hat{T}}_G$.

In both cases, the proof for the proposed algorithm will be structured as follows:
\begin{itemize}
\item First the so called \emph{geometric case} will be tackled: the vehicle is considered to always be at the desired heading angle, i.e. $\mathbf{\hat{T}}_M(t)=\frac{\mathbf{u}_{\text{fast,2}}}{k}(t), ~\forall t$.
\item Then, the so called \emph{dynamical case} (the vehicle is not always at the desired heading angle) will be considered and shown to fall into the geometrical case as time goes to infinity.
\end{itemize}

\subsection{Geometric case: finite paths}

\subsubsection{\textbf{Subcase 0. Single point path}}\label{Subcase 0: single point path}
Here we consider the path to be very far away and hence similar to a single point $P$ for the aircraft to be reached. 
The radially shifted distance is indistinguishable from the error, so $\theta_s(t)\approx 0~ \forall t$. 
Also, notice that with a point-path, $\mathbf{\hat{e}}=\mathbf{\hat{L}}_0$. 
By defining
\begin{equation}
a=\sqrt{{w^\star}^2-{v_M^\star}^2}, \qquad l=\|a\mathbf{\hat{L}}_0-\mathbf{w}\|
\end{equation}
%
we obtain
\begin{equation}
\begin{aligned}
\mathbf{v}_G\times \mathbf{\hat{L}}_0&=(w^\star+v_M^\star\mathbf{\hat{L}}_{1e})\times \mathbf{\hat{L}}_0=\\
&=(\mathbf{w}+\frac{v_M^\star(\mathbf{\hat{L}}_0a-\mathbf{w})}{l})\times \mathbf{\hat{L}}_0=\\
&=(\underbrace{(1-	\frac{v_M^\star}{l})}_{>0}\mathbf{w}+\frac{av_M^\star}{l}\mathbf{\hat{L}}_0)\times \mathbf{\hat{L}}_0\\
&=\underbrace{(1-	\frac{v_M^\star}{l})}_{>0}\mathbf{w} \times \mathbf{\hat{L}}_0 + 0
\end{aligned}
\end{equation}





Now let the line directed as $\mathbf{\hat{L}}_0$ divide the plane into two half-planes: the previous considerations imply that $\mathbf{v}_G$ and $\mathbf{w}$ both lie in the same half-plane, so the $\mathbf{\hat{L}}_0$ will rotate more towards the $-\mathbf{w}$ direction in time until eventually $\underset{t \rightarrow \infty}{\lim}\mathbf{\hat{L}}_0=\underset{t \rightarrow \infty}{\lim}\mathbf{\hat{e}}= -\mathbf{\hat{w}}$.
Another way to see this: the path-point $P$ acts as a rotational joint for the error vector $\mathbf{e}$, which is fixed at one end in $P$: the $\mathbf{v}_G$ is rotating the error vector in the same direction as the wind would rotate it, meaning that it will point instantaneously more in the anti-wind direction, i.e. even more outwardly with respect to the cone, until it reaches the antiwind direction (the ``torque'' around point $P$ is null at that point). \\
As in this case $\mathbf{\hat{L}}_0=\mathbf{\hat{e}}$, the  $\mathbf{\hat{L}}_0$ rotation must stop here.
Since $f(0)=0$, then also $\underset{t \rightarrow \infty}{\lim}\frac{\mathbf{u}_{\text{fast,2}}}{k} = -\mathbf{\hat{w}}$ by construction.
By hypothesis of geometrical case, this means $\underset{t \rightarrow \infty}{\lim}\mathbf{\hat{T}}_M(t)=-\mathbf{\hat{w}}$.
Then, by definition of the normal acceleration command, also $\underset{t \rightarrow \infty}{\lim}{\mathbf{a}_N^M} = 0$ so we reach \textbf{asymptotic safety} as defined in (\ref{eq:goals_finitelength}).


As an additional feature, note that
\begin{equation}
\begin{array}{l}
\text{sign}[(\mathbf{v}_G\times \mathbf{w}) \cdot \mathbf{k}](t_0)= \\
\quad \text{sign}[(\mathbf{v}_G\times \mathbf{w}) \cdot \mathbf{k}](t),~\forall t>t_0
\end{array}
\end{equation}
so we reach the equilibrium without oscillations around that line such that $\mathbf{\hat{e}}=-\mathbf{\hat{w}}$.


\subsubsection{\textbf{Subcase 1. Finite length paths}}

In this case, the $\mathbf{\hat{L}}_0$ versor is a  function of the  particular path we are considering, so we can no longer assume it to coincide with $\mathbf{\hat{e}}$ as in the single point path case.
%


\noindent However, consider the following two facts:
\begin{itemize}
\item The path is finite
\item $\forall t\in \mathbb{R}, ~~\mathbf{r}_M(t) \cdot \mathbf{\hat{w}}~\geq~\mathbf{r}_M(0)\cdot \mathbf{\hat{w}}+\underbrace{(w^\star-v_M^\star)}_{>0}t $
\end{itemize}

That is, as the wind is constantly stronger than the airspeed, the minimum growth of the projection of the error onto the wind direction has rate $(w^\star-v_M^\star)t$.
This implies that
\begin{equation}
\lim_{t\rightarrow +\infty}\|\mathbf{r}_M(t)\|=+\infty
\end{equation}
and since the path is finite
\begin{equation}
\lim_{t\rightarrow +\infty}\|\mathbf{d}(t)\|=\lim_{t\rightarrow +\infty}\|\mathbf{e}(t)\|=+\infty
\end{equation}

As the distance grows to infinity, the path will look like a single point $\mathbf{P}_\infty$, that is the center of the smallest circle that contains the whole path. 
\noindent Then, we fall into the single point-path subcase.

\subsection{Geometric case: infinite linear paths}\label{Subcase 1: infinite linear paths}
Here the path is not finite. However, a common case in UAV applications is when the path is an infinite line. If this line is outside the wind-cone or the intersection with the cone is finite, it is not possible for the $\mathbf{v}_G$ to align to it. In this case, the proposed algorithm achieves \textbf{efficient wind stability}, i.e. the objectives in (\ref{eq:goals_linear}).
To show this, simply notice that if $\mathbf{d}>\delta_{BL}$, then whatever the vehicle position, we have ${\mathbf{\hat{L}}_0}~ \mathbf{\bot}~ \hat{\Lambda}$, as the error direction will always be perpendicular to the line. A simulation for this situation is shown in Figure \ref{partial_wind_stability}.


\noindent The interpretation for this result is that the proposed algorithm finds some efficient compromise for the $\mathbf{v}_G$ direction between the anti-wind direction and the path direction, which is a tradeoff between safety and tracking performance.

\begin{figure}[thpb]
      \centering
      \includegraphics[width=0.95\columnwidth]{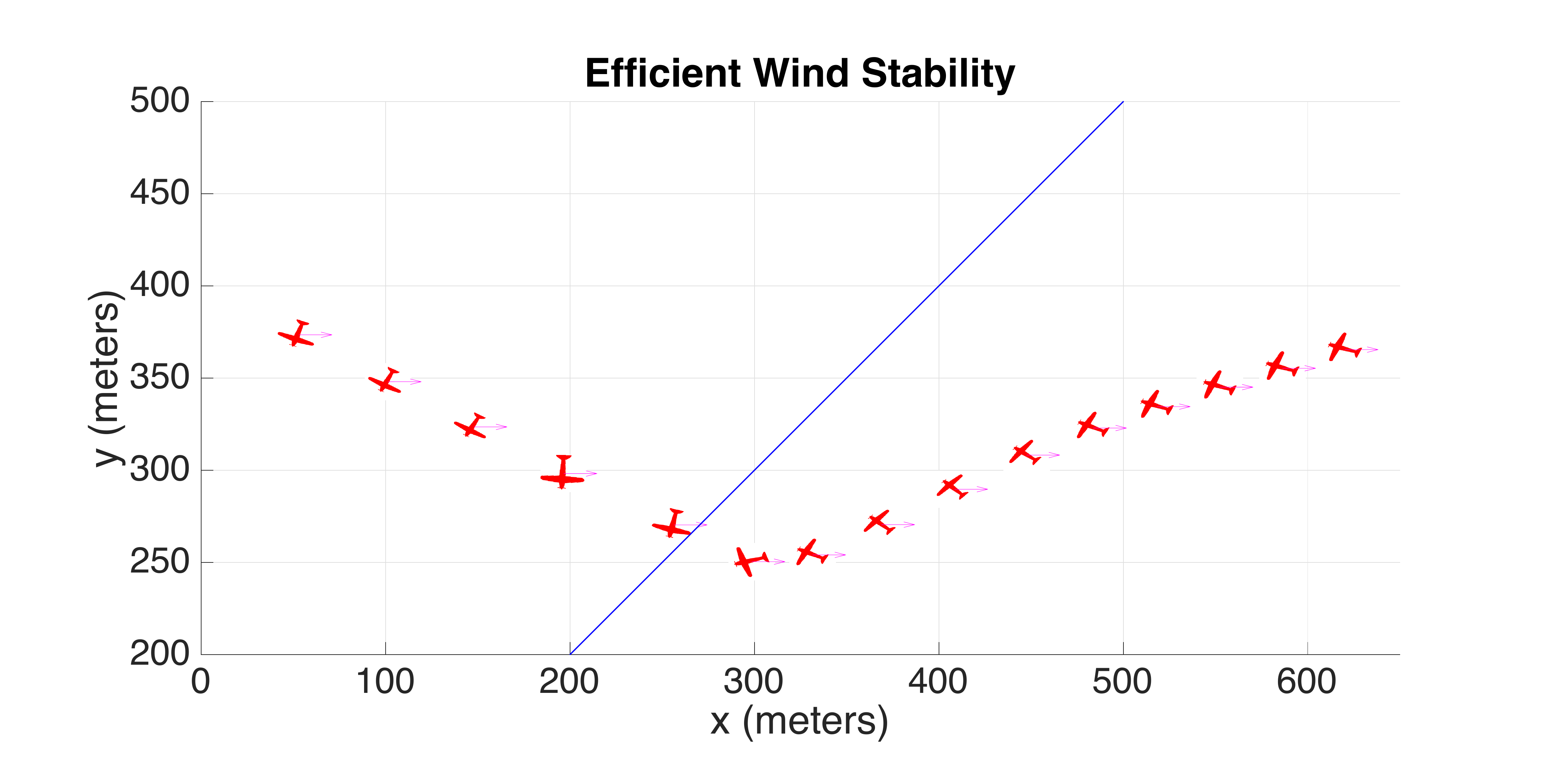}
      \caption{Airspeed is 14 m/s. Windspeed is 30 m/s, indicated by the magenta arrow on the aircraft. $\hat{\Lambda}=(\frac{\sqrt{2}}{2},\frac{\sqrt{2}}{2})$. Line direction is hence infeasible.}
\label{partial_wind_stability}
\end{figure}







\subsection{Dynamical case}
Here we will extend the proof for the geometric case, so as to consider the dynamics imposed by the nonlinear acceleration command. As the subcase of finite-length paths was shown to fall into the subcase of single-point paths, studying the dynamic extension for the single-point paths is all we need. The extension for the infinite linear path is trivial and will be omitted, as the $\mathbf{\hat{L}}_0$ stops changing as soon as $\mathbf{d}>\delta_{BL}$.



\noindent In the following, it is clearer to directly refer to Figure (\ref{proof_notation}) for the symbols definition. 
\begin{figure}[thb]
      \centering
      \includegraphics[width=0.95\columnwidth]{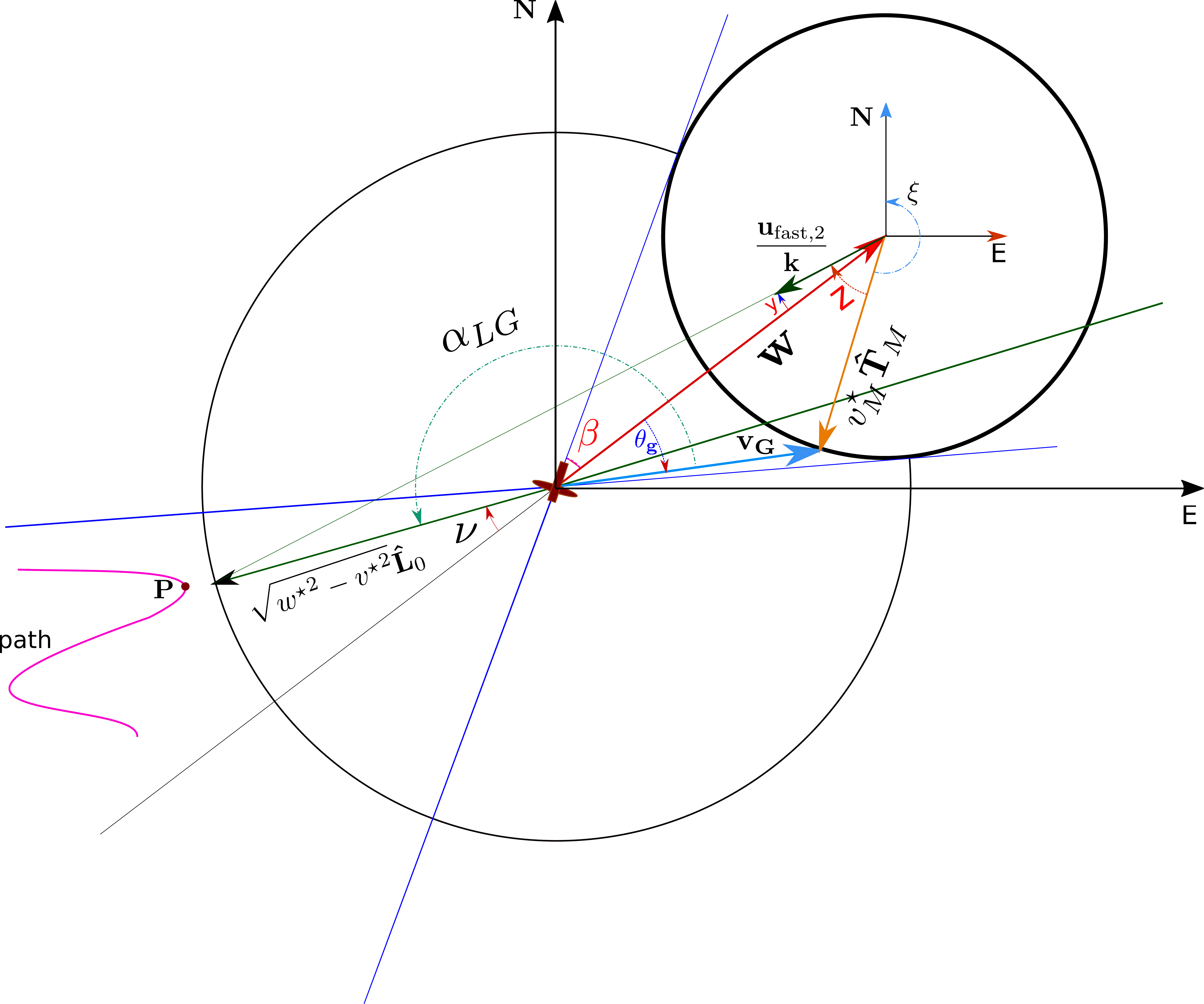}
      \caption{Symbols used in the proof}
\label{proof_notation}
\end{figure}
%

\noindent Depending on the desired groundspeed direction $\mathbf{\hat{L}}_0$, we have two subcases.
\subsubsection{Subcase 1}
\noindent If
\begin{equation}
\beta<\nu<\pi-\beta
\end{equation}
\noindent corrensponding to $\mathbf{\hat{L}}_0$ pointing outside of the ``specular'' cone, then it is easy to see that 
\begin{equation}
\alpha_{Lg}<\pi,~\forall \theta_g
\end{equation}
\noindent meaning that
\begin{equation}
\dot{\nu}<0
\end{equation}
\noindent independently from the actual aircraft orientation. This holds until we fall into subcase 2.
\subsubsection{Subcase 2}
If
\begin{equation}
0<\nu<\beta
\end{equation}
corrensponding to $\mathbf{\hat{L}}_0$ pointing inside of the ``specular'' cone, we need further considerations. It is not true anymore that $\alpha_{Lg}<\pi,~\forall \theta_g$. Instead we have

\begin{equation}
\begin{cases}
\alpha_{Lg}<\pi,~\text{if }\theta_g>-\nu\\
\alpha_{Lg}\geq \pi \text{ otherwise}
\end{cases}
\end{equation}
\noindent Then it's possible that, depending on how the aircraft is oriented, $\nu$ will increase while the angle between $\mathbf{\hat{T}}_M$ and the commanded direction $\frac{\mathbf{u}_{\text{fast,2}}}{k}$ is smaller than $\pi$ , which is undesirable as it would mean the $\mathbf{\hat{L}}_0$ is ``running away'' from $\mathbf{\hat{T}}_M$.

	\noindent To show that eventually the aircraft can be considered to be aligned with its commanded control input versor $\frac{\mathbf{u}_{\text{fast,2}}}{k}$, consider the following:
\begin{itemize}
\item We can increase parameter $k$ in order to make the vehicle turn with  faster dynamics.
\item As time goes to infinity, eventually the ``chasing'' angle $z$ will decrease to 0.
\end{itemize}

To show this last fact, first notice that for any given $\mathbf{\hat{T}}_M$, if $\dot{\nu}>0$ then $\dot{\nu}$ is a decreasing function of $|\mathbf{e}\cdot \mathbf{w}|$ that goes to 0 as $\frac{1}{|\mathbf{e}\cdot \mathbf{w}|}$ or faster. Indeed, consider the case when $\dot{\nu}>0$ and has the maximum value, i.e. $\nu=0$ and $\hat{T}_M \mathbf{\bot} \mathbf{w}$. We have
\begin{equation}
\dot{\nu}_{\text{MAX}}=\frac{v_M^\star}{|\mathbf{e}\cdot \mathbf{w}|}
\end{equation}
which acts as an upperbound for all the other situations. Irrespectively from $\mathbf{\hat{T}}_M$, since $w^\star>v_M^\star$, $|\mathbf{e}\cdot \mathbf{w}|$ indeed increases, hence $\dot{\nu}$ must decrease and tend to 0.
Since $y$ is a function of $\nu$ such that $\forall~ \nu,~y(\nu)<\nu$, than also $\dot{y}$ decreases and tends to 0 as time goes to infinity.
\noindent Now consider the time derivative of the ``chasing'' angle $z$
\begin{equation}
\dot{z}=\dot{y}+\dot{\xi}
\end{equation}




Since we showed $\underset{t\rightarrow +\infty}{\lim}\dot{\nu}(t)=0=\underset{t\rightarrow +\infty}{\lim}\dot{y}(\nu(t))$, irrespectively of what the orientation of the vehicle could be at any time, then, as $\xi$ indicates the heading angle of the aircraft,
\begin{equation}
\label{eq:z infinity}
\underset{t\rightarrow \infty}{\lim}\dot{z}(t)=\dot{\xi}(t)
\end{equation}
%
\noindent As the acceleration command is designed to steer the vehicle orientation onto the chosen look-ahead vector, which now is $\frac{\mathbf{u}_{\text{fast,2}}}{k}$, as the look-ahead is bound to asymptotically stop changing as the vehicle gets further away from the path, then we actually have that  $\underset{t \rightarrow \infty}{\lim}\dot{\xi}(t) = 0$, with the vehicle aligned to the look-ahead vector. This, together with  (\ref{eq:z infinity}), translates in

\begin{equation}
\underset{t \rightarrow \infty}{\lim}z(t)=0
\end{equation}
%
\noindent Then we can say that we asymptotically fall into the geometrical case, and the proof holds.
} 


\bibliographystyle{IEEEtran}
\bibliography{references.bib}

\end{document}